\documentclass[12pt]{article}
\pdfoutput=1 

\usepackage{cancel,slashed}
\usepackage{bbm}
\usepackage{amssymb}
\usepackage{amsfonts}
\usepackage{amsmath}
\usepackage{graphicx}
\usepackage{cite}

\usepackage{latexsym}
\usepackage{color}
\usepackage{xypic}
\usepackage{cancel,slashed}
\usepackage[linktocpage]{hyperref}
\usepackage{color}
\usepackage{transparent}
\usepackage{footmisc}

\makeatletter
\def\@xfootnote[#1]{%
  \protected@xdef\@thefnmark{#1}%
  \@footnotemark\@footnotetext}
\makeatother

\newcommand{\bmat}{\left(\begin{array}}
\newcommand{\emat}{\end{array}\right)}

\def\vec#1{\textbf{\emph{#1}}}

\def\-{\hphantom{-}}

\def\s2{\frac{1}{\sqrt2}}

\def\beq{\begin{equation}}
\def\eeq{\end{equation}}
\def\beqa{\begin{eqnarray}}
\def\eeqa{\end{eqnarray}}

\def\im{{\rm Im \,}}
\def\re{{\rm Re \,}}

\def\Z{{\mathbb Z}}

\def\Dsl{\,\raise.15ex\hbox{/}\mkern-13.5mu D} 

\def\re{\mbox{Re}}
\def\im{\mbox{Im}}

\def\be{\begin{equation}}
\def\ee{\end{equation}}
\def\bea{\begin{eqnarray}}
\def\eea{\end{eqnarray}}
\def\raw{\rightarrow}

\def\IN{\mathbb{N}}
\def\IZ{\mathbb{Z}}

\def\Lam{{\Lambda}}

\def\sig{{\sigma}}

\def\vec#1{{\overrightarrow{#1}}}

\makeatletter
\newsavebox{\@brx}
\newcommand{\llangle}[1][]{\savebox{\@brx}{\(\m@th{#1\langle}\)}%
  \mathopen{\copy\@brx\kern-0.5\wd\@brx\usebox{\@brx}}}
\newcommand{\rrangle}[1][]{\savebox{\@brx}{\(\m@th{#1\rangle}\)}%
  \mathclose{\copy\@brx\kern-0.5\wd\@brx\usebox{\@brx}}}
\makeatother

\def\sm2{{\mbox{\small 2}}}

\newcommand{\bp}{\begin{pmatrix*}[r]}  
\newcommand{\ep}{\end{pmatrix*}}  
\newcommand{\bpp}{\begin{pmatrix}}  
\newcommand{\epp}{\end{pmatrix}}  
\newcommand{\bcd}{\begin{center}
\begin{tikzcd}}
\newcommand{\ecd}{\end{tikzcd} \end{center}}

\def\1{\mathbb{1}}

\newenvironment{eqn}{\begin{equation}\begin{aligned}}{\end{aligned}\end{equation}\noindent}
\newenvironment{eqn*}{\begin{equation*}\begin{aligned}}{\end{aligned}\end{equation*}\noindent}

\begin{document}
\pagestyle{plain}

\makeatletter
\@addtoreset{equation}{section}
\makeatother
\renewcommand{\theequation}{\thesection.\arabic{equation}}
\pagestyle{empty}
\rightline{ IFT-UAM/CSIC-19-049}
\vspace{0.5cm}
\begin{center}
\Huge{{Instantons and infinite distances}
\\[15mm]}
\normalsize{Fernando Marchesano$^1$ and Max Wiesner$^{1,2}$  \\[10mm]}
\small{
${}^1$Instituto de F\'{\i}sica Te\'orica UAM-CSIC, Cantoblanco, 28049 Madrid, Spain \\[2mm] 
${}^2$ Departamento de F\'{\i}sica Te\'orica, 
Universidad Aut\'onoma de Madrid, 
28049 Madrid, Spain
\\[8mm]} 
\small{\bf Abstract} \\[5mm]
\end{center}
\begin{center}
\begin{minipage}[h]{15.0cm} 

We consider geodesics of infinite length and constant 4d dilaton in the (classical) hypermultiplet moduli space of type II Calabi-Yau compactifications. When approaching such infinite distance points, a large amount of D-instantons develop an exponentially suppressed action, substantially modifying the moduli space metric. We consider a particular large volume/strong coupling trajectory for which, in the corrected metric, the path length becomes finite. The instanton effects also modify the cllassical 4d dilaton such that, in order to keep the 4d Planck mass finite, the string scale has to be lowered. Our results can be related, via the c-map, to the physics around points of infinite distance in the vector multiplet moduli space where the Swampland Distance Conjecture and the Emergence Proposal have been discussed, and provide further evidence for them. 

\end{minipage}
\end{center}
\newpage
\setcounter{page}{1}
\pagestyle{plain}
\renewcommand{\thefootnote}{\arabic{footnote}}
\setcounter{footnote}{0}


\tableofcontents


\section{Introduction}
\label{s:intro}

The Swampland Program \cite{Vafa:2005ui} (see \cite{Brennan:2017rbf,Palti:2019pca} for reviews) addresses one of the most fascinating questions in modern theoretical High Energy Physics: the constraints that quantum gravity imposes on effective field theories. It also lies at the core of one of the most important aspects of String Theory: its predictivity. A large fraction of the recent activity in this topic stems from the initial swampland conjectures \cite{ArkaniHamed:2006dz,Ooguri:2006in}. From these, particular attention has recently been given to the Swampland Distance Conjecture (SDC), which states that the field space of the effective theory contains geodesic paths of infinite distance and that, when reaching their endpoints, an infinite tower of resonances decrease their mass exponentially fast. The analysis of this conjecture in different string theory setups \cite{Baume:2016psm,Klaewer:2016kiy,Valenzuela:2016yny,Blumenhagen:2017cxt,Palti:2017elp,Hebecker:2017lxm,Grimm:2018ohb,Heidenreich:2018kpg,Blumenhagen:2018nts,Landete:2018kqf,Lee:2018urn,Reece:2018zvv,Lee:2018spm,Ooguri:2018wrx,Grimm:2018cpv,Buratti:2018xjt,Hebecker:2018fln,Gonzalo:2018guu,Corvilain:2018lgw,Lee:2019tst,Blumenhagen:2019qcg,Joshi:2019nzi,tensionless}, has eventually led to support the Emergence Proposal \cite{Harlow:2015lma,Grimm:2018ohb,Heidenreich:2017sim,Heidenreich:2018kpg,Palti:2019pca}, in which weak gauge couplings and infinite distances in field space are postulated to arise from the integration of a tower of light fields. 

Besides trying to understand the different proposals within the Swampland Program at a fundamental level, one would also like to test them in the current set of string theory vacua \cite{thebook,Baumann:2014nda}. To do so, one must understand the interplay of the different Swampland Conjectures with the ingredients in such string constructions. Two ingredients of this sort are internal background fluxes and non-perturbative effects, which play a key role in the proposals for de Sitter vacua populating the Landscape \cite{Kachru:2003aw,Balasubramanian:2005zx}. On the one hand, potentials generated by fluxes do play a role in Swampland-related ideas challenging such proposals \cite{Obied:2018sgi,Garg:2018reu}, which have been connected to the SDC and the Emergence Proposal in \cite{Ooguri:2018wrx,Palti:2019pca}. On the other hand, the role of non-perturbative effects remains less clear in this picture. In fact, the interplay between the Swampland Conjectures and non-perturbative effects has mostly originated from the Weak Gravity Conjecture (WGC) for axions \cite{ArkaniHamed:2006dz}, which constrains models of natural inflation and generalisations, see e.g. \cite{Rudelius:2015xta,Montero:2015ofa,Brown:2015lia}.

The purpose of this work is to confront non-perturbative effects with yet another Swampland Conjecture, namely the Swampland Distance Conjecture. Notice that, a priori, the WGC for axions does not prevent trans-Planckian periodicities. It only states that in this case non-perturbative effects will be strong enough to spoil the one-instanton approximation. This is a serious drawback when trying to generate monotonic potentials via non-perturbative effects, but it may not be an issue in other instances. In particular, one may consider models with extended supersymmetry like 4d ${\cal N} = 2$ theories, where the effect of instantons is to modify the metric for the moduli space of hypermultiplets, but no potential is generated. 

A well-known set of 4d ${\cal N} = 2$ theories is obtained from compactifying type II string theory on Calabi-Yau manifolds. This class of compactifications has been extensively studied in the literature, and has led to remarkable results like the resolution of the conifold singularity in the vector multiplet moduli space by taking into account the presence of a light D-particle \cite{Strominger:1995cz}, and its counterpart in the hypermultiplet moduli space by a large number of D-instantons \cite{Ooguri:1996me}. The relation between these two effects can be made precise by means of the c-map \cite{Cecotti:1988qn,Ferrara:1989ik}, which maps D-particles charged under the vector multiplets of type IIA/B string theory on $X$ to D-instantons modifying the hypermultiplet moduli space of type IIB/A on the same Calabi-Yau. Interestingly, the same underlying principle that allows to resolve the conifold singularity in \cite{Strominger:1995cz} was invoked in \cite{Grimm:2018ohb} to propose the emergence of infinite distances and weakly coupled gauge interactions in CY vector moduli spaces. The main difference with respect to the case in \cite{Strominger:1995cz} is that, instead of one, an infinite set of D-particles becomes light when reaching an infinite distance point, in agreement with the Swampland Distance Conjecture.

In this paper we analyse the behaviour of the ${\cal N}=2$ theory along trajectories in the hypermultiplet moduli space. This is in general a complicated problem because, unlike its vector multiplet counterpart, the classical metric for hypermultiplets receives  $g_s$ and $\alpha'$-corrections, both at the perturbative and non-perturbative level. Nevertheless, we consider a region in moduli space in which the quantum corrected metric can be computed. In this region one may define infinite distance geodesics in terms of the classical moduli space metric, before any perturbative or non-perturbative corrections have been taken into account, and then see what is the effect of the quantum corrections. We find that, as we approach certain (classical) infinite distance point, towers of towers of D-instantons decrease their action very fast -- exponentially fast in the classical proper distance -- substantially modifying the metric. The modification is such that the infinite distance geodesic is no longer so in the quantum corrected metric. 

Besides the metric, the presence of small-action instantons modify the classical relation between the 4d Planck mass and the string scale. Near the classical infinite distance point the ratio $M_{\rm P}/M_s$ blows up, and goes like $\sqrt{N}$, where $N$ can be understood as the number of D-instantons towers that contribute non-trivially to the metric. Therefore, if one insists to keep the 4d Planck mass constant the string scale has to be lowered accordingly, acting like a species scale. If on the contrary one keeps $M_s$ fixed, these become points in which gravity inevitably decouples. 

These two results are very suggestive from the viewpoint of the setup in \cite{Grimm:2018ohb}, which can be related to ours by compactifying on a circle and applying a chain of dualities. Indeed, they can both be related to one-loop corrections involving $N$ light D-particles in the dual theory. In particular, the quantum corrected metric can be understood as generated from one-loop corrections as in the original Emergence proposal, but now in the case of a three-dimensional effective theory. This can in turn be interpreted as an interesting extension of the Swampland Distance Conjecture to the three-dimensional case. Moreover, it shows that the use of instanton-corrected metrics constitutes a systematic and precise method to encode the physics of many light D-particles, and therefore a powerful tool to test Swampland criteria.

The paper is organised as follows. In Section \ref{s:typeIIB} we describe the setup in which we will perform our computations, namely type IIB string theory on a Calabi-Yau, and the classical infinite distance trajectory that we consider. The main tool for our analysis will be the so-called tensor potential, which automatically encodes the relevant D-instanton effects. In Section \ref{s:towers} we will analyse the behaviour of this tensor potential along the said trajectory, together with that of the quantum corrected metric that can be derived from it. We compute the trajectory length in the quantum corrected metric, finding it to be finite, and the quantum corrected kinetic terms for the periodic directions, which we interpret in terms of the electric and magnetic WGC for axions. Finally, we interpret our findings in terms of the dual setup with D-particles in three-dimensions, and we draw our conclusions in Section \ref{s:conclu}. 

Several technical details have been relegated to the appendices. In Appendix \ref{ap:cmap} we review the c-map relating vector multiplet and hypermultiplet moduli spaces. In Appendix \ref{ap:conifold} we discuss how the contact potential can be used to resolve conifold singularities in the hypermultiplet moduli space of type IIB CY compactifications. In Appendix \ref{ap:metric} we detail the prescription used to compute the asymptotic behaviour of the type IIB hypermultiplet moduli space metric for the region of interest in the main text. Finally, Appendix \ref{ap:typeIIA} extends our discussion to the case of hypermultiplet moduli spaces in type IIA compactifications, where we compute the exact metric using the standard methods in the literature.

\section{Type IIB on Calabi-Yau manifolds and instantons}
\label{s:typeIIB}

In this section we describe the framework in which we perform our analysis. Throughout most of the paper we  consider type IIB string theory compactified on a Calabi-Yau. The instantons that appear in the 4d ${\cal N}=2$ effective theory will affect the metric of the hypermultiplet moduli space. As we will discuss, along a trajectory of infinite distance which increases the volume and the string coupling, such non-perturbative effects can be encoded in a function dubbed tensor potential, which we estimate. In the next section we analyse the consequences of such corrections for the hypermultiplet moduli space metric.

\subsection{The type IIB hypermultiplet moduli space}
\label{ss:IIBHM}

Let us consider type IIB string theory compactified on a Calabi-Yau three-fold $X$. At low energies, one obtains a 4d ${\cal N} =2$ effective theory whose field content is given by a gravity multiplet, $h^{2,1}(X)$ vector multiplets and $h^{1,1}(X)+1$ hypermultiplets. The scalars within these multiplets parametrise the moduli space of the effective theory which, by general  4d  ${\cal N} =2$ arguments, factorises at the two-derivative level as
\begin{equation}
{\cal M} = {\cal M}_{\rm VM} \times {\cal M}_{\rm HM}\, .
\end{equation}
Here ${\cal M}_{\rm VM}$ is a special K\"ahler manifold described by $h^{2,1}$ vector multiplet complex scalars \cite{deWit:1984wbb}, and ${\cal M}_{\rm HM}$ is a quaternionic-K\"ahler manifold parametrised by $4(h^{1,1}+1)$ real scalars within the hypermultiplets \cite{Bagger:1983tt}. In the following we will be mostly interested in the hypermultiplet sector of the theory, whose scalar field content is summarised in table \ref{hypers}: 
\begin{table}[h]
\begin{center}
\begin{tabular}{llll}
universal hypermultiplet & & $\tau = C_0 + i e^{-\phi}$, & $b^0$, \quad $c^0$\\
$h^{1,1}$ hypermultiplets & & $z^a  = b^a + it^a$, & $c^a$ ,\quad  $d^a$
\end{tabular}
\caption{Type IIB hypermultiplet scalar content, whose vevs  are the coordinates of ${\cal M}_{\rm HM}$.}
\label{hypers}
\end{center}
\end{table}

\vspace*{-.5cm}

 The universal hypermultiplet contains the axion dilaton $\tau = \tau_1  + i \tau_2 =  C_0 + i e^{-\phi}$ and the scalar duals $b^0$, $c^0$ of the 4d two-forms coming from the B-field $B$ and the RR potential $C_2$, respectively. The remaining $h^{1,1}$ hypermultiplets contain the complexified K\"ahler coordinates of the Calabi-Yau
 \begin{equation}
 z^a  = b^a + it^a =  \ell_s^{-2} \int_{\gamma^a} B +i J\, ,
 \end{equation}where $\ell_s = 2\pi \sqrt{\alpha'}$ is the string length, and $\{\gamma^a\}$ is a basis of $H_2(X,\Z)$ such that all areas $t^a$ are positive. They also contain the integrals of the RR potentials over such two-cycles 
 \begin{equation}
 c^a = \ell_s^{-2}\int_{\gamma^a} C_2\, , \qquad \qquad  d_2^a = \ell_s^{-2}\int_{\gamma^a} C_4\, ,
 \end{equation}where the two-forms $d_2^a$ are the 4d duals of the scalars $d^a$.

Describing the type IIB hypermultiplet moduli space metric is in general seen as an arduous problem because, unlike its vector multiplet counterpart, its classical expression receives all kinds of  $g_s$ and $\alpha'$-corrections, both at the perturbative and non-perturbative level. Particularly difficult to handle are the non-perturbative corrections that that arise from D-branes and NS5-branes wrapping even-dimensional cycles in the internal manifold $X$, and seen as instantons in the 4d effective theory \cite{Becker:1995kb,Marino:1999af}. Nevertheless,  a remarkable amount of progress has been achieved in this direction, by a combined use of dualities, discrete symmetries and twistor methods  \cite{Rocek:2005ij,RoblesLlana:2006ez,RoblesLlana:2006is,Neitzke:2007ke,RoblesLlana:2007ae,Alexandrov:2008gh,Alexandrov:2009zh,Alexandrov:2010ca} (see \cite{Alexandrov:2011va} for a review). One key ingredient in this approach is the transformation known as c-map \cite{Cecotti:1988qn,Ferrara:1989ik}, which we review in Appendix \ref{ap:cmap}. The c-map relates the vector moduli space of type IIA/IIB string theory compactified on $X$ with the hypermultiplet moduli space of type IIB/IIA compactified on the same Calabi-Yau, embedding ${\cal M}_{\rm VM}^{\rm IIA/IIB}(X)$ into ${\cal M}_{\rm HM}^{\rm IIB/IIA}(X)$ as a totally geodesic manifold \cite{Cecotti:2015wqa}. Moreover, it maps IIA/IIB 4d D-particles to type IIB/IIA D-instantons wrapping the same internal cycles of $X$. 

The properties of the c-map are very suggestive when combined with the results of \cite{Grimm:2018ohb,Grimm:2018cpv,Corvilain:2018lgw}, and allow to analyse them from a different perspective. Indeed, in the setup in \cite{Grimm:2018ohb,Grimm:2018cpv,Corvilain:2018lgw} infinite towers of D-particles become massless exponentially fast when approaching points of infinite distance along geodesics of ${\cal M}_{\rm VM}$, 
giving a neat realisation of the Swampland Distance Conjecture \cite{Ooguri:2006in}. When embedding such geodesics into ${\cal M}_{\rm HM}$ via the c-map, D-particles should be replaced by D-instantons with exponentially decreasing action. The importance of such D-instanton effects on the moduli space metric should parallel the relevance of the light D-particles one-loop contribution, whose precise form is crucial to the Emergence Proposal. In the following we will see that, indeed, along paths of infinite distance in ${\cal M}_{\rm HM}^{\rm IIB}(X)$ infinite towers of D-instantons develop an exponentially suppressed action, dramatically modifying the classical moduli space metric.

\subsection{Geodesics of infinite distance}
\label{ss:geodesic}

Within the scalar field content of table \ref{hypers}, only $\tau_2 = e^{-\phi}$ and the $t^a$ are non-periodic coordinates of ${\cal M}_{\rm HM}^{\rm IIB}(X)$. It is thus natural to construct geodesics of infinite distance by performing rescalings on such fields. One may in particular consider rescalings increasing the value of the K\"ahler coordinates $t^a$,  corresponding to different decompactification limits. As discussed in \cite{Corvilain:2018lgw,tensionless},  the monodromy orbits around the corresponding points of infinite distance are well understood and can be classified. 

In the following we will consider the following trajectory in ${\cal M}_{\rm HM}^{\rm IIB}(X)$
\begin{equation}
t^a (\sigma) = e^\sigma t^a (x_0)\, , \qquad \qquad \tau_2(\sigma) = e^{-\frac{3}{2}\sigma} \tau_2(x_0)\, ,
\label{limit}
\end{equation}
where $\sigma \in (0, \infty)$ parametrises the trajectory and $x_0$ is an interior point of ${\cal M}_{\rm HM}^{\rm IIB}(X)$. For simplicity we will consider $x_0$ such that the periodic fields in table \ref{hypers} have vanishing vev. One can then easily show that, with the classical metric, the above path corresponds to a geodesic of infinite length, as we discuss in the next section. This choice also allows to use the c-map to describe a trajectory in ${\cal M}_{\rm VM}^{\rm IIA}(X)$, where the periodic fields $(b^0, c^0, c^a, d^a)$ are absent. In this sense, notice that the trajectory \eqref{limit} is different form the ones taken in \cite{Corvilain:2018lgw} because it not only takes us to large volume, but also to strong coupling. The reason for taking this limit is to keep the classical 4d Planck mass invariant
\begin{equation}
M_{\rm P} =  \tau_2  \sqrt{2\pi V(t)}\, M_s\, ,
\end{equation}
where $M_s = \ell_s^{-1}$ is the string scale, and $V(t) = \frac{1}{6} {\cal K}_{abc}t^at^bt^c$ stands for the volume of the Calabi-Yau manifold $X$ in string units and in the string frame, with ${\cal K}_{abc}$ the triple intersection numbers of $X$. This kind of field space trajectory keeping the Planck mass constant has also been considered in \cite{tensionless}, among other possibilities.

In \cite{Corvilain:2018lgw} it was argued that, even if the large volume limit takes $M_P \raw\infty$, one may still consider quotients of the form $m/M_{\rm P}$, where $m$ is the mass of a certain mode, and that such quotients may decrease exponentially near the infinite distance point. It is easy to see that if $m$ corresponds to, e.g. a D-particle, the dilaton dependence will cancel out, and the mass ratio along the path will not depend on whether we rescale $\tau_2$ or not. However, the same kind of reasoning cannot be applied to D-instantons, because in this case there is no scale to which to compare their individual action. We will therefore stick to the dilaton-dependent trajectory \eqref{limit} which, in the original spirit of the Swampland Program \cite{Vafa:2005ui,Ooguri:2006in}, in principle avoids gravity-decoupling limits.

In fact, following the trajectory \eqref{limit} in ${\cal M}_{\rm VM}^{\rm IIA}(X)$ is mirror symmetric to the kind of infinite distance directions studied in \cite{Grimm:2018ohb} for ${\cal M}_{\rm VM}^{\rm IIB}(Y)$, where $Y$ is the Calabi-Yau manifold mirror to $X$. Indeed, for such a Calabi-Yau manifold we have that
\begin{equation}
V_Y 
= \frac{i}{8} \int_Y \Omega \wedge \overline{\Omega} \simeq |X^0|^2 \frac{1}{6} {\cal K}_{abc} \zeta^a\zeta^b\zeta^c
\end{equation}
where in the last step we have approximated $\Omega$ by its large complex structure expression. Here $\zeta^a = \im (X^a/X^0)$ are defined in terms of the periods $\int_{\Pi^a}\Omega$ over the $A$-cycles, and in particular $X^0$ stands for the period of the reference three-cycle $\Pi^0$. In the large complex structure limit $\zeta^a \raw \infty$, one can only keep $V_Y$ finite by taking $|X^0|\raw 0$, collapsing the reference $A$-cycle $\Pi^0$, whose volume plays the same role as $\tau_2$ in \eqref{limit}. In fact, upon mirror symmetry a D3-brane wrapping $X^0 \subset Y$ becomes a D0-brane pointlike in $X$, and $g_{s,\,\rm IIB}^{-1}|X^0|$ is mapped to $g_{s,\,\rm IIA}^{-1}$. Therefore, a decreasing volume of the reference $A$-cycle in $Y$ maps to a strong coupling limit on its mirror manifold $X$.\footnote{Conversely, the mirror  trajectory to \eqref{limit} in ${\cal M}_{\rm HM}^{\rm IIA}(Y)$ does not involve any strong coupling regime. As discussed in Appendix \ref{ap:typeIIA}, all our results can also be recovered in this type IIA framework.}

Just like in \cite{Corvilain:2018lgw,tensionless}, one may consider different geodesic trajectories of infinite distance by applying different scalings for the K\"ahler coordinates $t^a$. Rather than analysing such different possibilities, we will focus on the simple case \eqref{limit} where all the K\"ahler coordinates are treated universally and so are the 4d instantons. Indeed, one can easily check that the actions of the different instantons scale as follows:
\begin{table}[h]
\begin{center}
\begin{tabular}{lll}
$S_{D(-1)}$ & $\raw$ & $e^{-\frac{3}{2}\sigma} S_{D(-1)}$ \\
$S_{D1}$ & $\raw$ & $e^{-\frac{1}{2}\sigma} S_{D1}$\\
$S_{D3}$ & $\raw$ & $e^{\frac{1}{2}\sigma} S_{D3}$ \\
$S_{D5}$ & $\raw$ & $e^{\frac{3}{2}\sigma} S_{D5}$ \\
$S_{F1}$ & $\raw$ & $e^{\sigma} S_{F1}$  \\
$S_{NS5}$ & $\raw$ & $S_{NS5}$\, .
\end{tabular}
\end{center}
\end{table}

\vspace*{-.5cm}

Therefore, for large values of $\sigma$, the relevant non-perturbative corrections amount to D(-1) and D1-brane instantons or bound states of them. In the following we will analyse their effect, together with those of perturbative corrections, on the metric of ${\cal M}_{\rm HM}^{\rm IIB}(X)$.

\subsection{The tensor potential at infinite distance}
\label{ss:Kinfinite}

The metric of the quaternionic-K\"ahler manifold ${\cal M}_{\rm HM}$ can be encoded in a real function $\chi$, which essentially plays the role of a K\"ahler potential \cite{deWit:1999fp,deWit:2001brd}. In the type IIB case, whenever D3, D5 and NS5 instantons can be neglected, the hypermultiplet sector can be described in terms of tensor multiplets, and $\chi$ can be computed in terms of a potential for the latter \cite{deWit:2006gn}. Such a potential, usually dubbed tensor or contact potential, has been obtained in \cite{RoblesLlana:2006is} and can be seen as a sum of two terms
\begin{equation}
\chi = \chi_{\rm cl} + \chi_{\rm corr}\, ,
\label{contact}
\end{equation}
where
\begin{equation}
\chi_{\rm cl} = \frac{1}{12} \tau_2^2  {\cal K}_{abc}t^at^bt^c \, , 
\label{chiclas}
\end{equation}
is the classical contribution to the potential and
\begin{equation}
\chi_{\rm corr} = \frac{\tau_2^2}{8(2\pi)^3} \sum_{\bf{k}\geq 0} n_{\bf{k}}^{(0)} \sum_{(m,n)\in \mathbb{Z}^2\backslash 0}  \frac{1+ 2\pi |m\tau + n|k_at^a}{|m\tau + n|^3}\, e^{-S_{m,n}^{\bf{k}} }\, ,
\label{chicorr}
\end{equation}
are the relevant perturbative and non-perturbative corrections. Here {\bf k} is a vector of $h^{1,1}$ entries $k_a \in \IN$, such that $k_a \gamma^a$ scans the homology classes in $H_2^+(X,\mathbb{Z})$. The terms with $\bf{k}\neq 0$ represent the corrections coming from Euclidean $(m,n)$-strings wrapping two-cycle classes with non-vanishing genus-zero Gopakumar-Vafa invariant $n_{\bf{k}}^{(0)}$, with action
\begin{equation}
S_{m,n}^{\bf{k}} = 2\pi k_a \left(|m\tau + n| t^a  - imc^a - inb^a \right)\, .
\label{Smn}
\end{equation}
The contribution from the term ${\bf k} = 0$ corresponds to the D(-1)-brane corrections, together with the one-loop corrections on $g_s$ and $\alpha'$, if one sets $n_{\bf{k}=0}^{(0)}= - \chi_E(X)$, that is to (minus) the Euler characteristic of $X$. 

In general, the metric in ${\cal M}_{\rm HM}^{\rm IIB}(X)$ will receive further corrections not captured by $\chi_{\rm corr}$, coming from Euclidean D3, D5 and NS5-branes. These will be however negligible in certain limits, like when resolving  conifold singularities via instantons \cite{Ooguri:1996me}. Indeed, as shown in \cite{Saueressig:2007dr} and in Appendix \ref{ap:conifold}, one may use \eqref{chicorr} to see how a conifold singularity in ${\cal M}_{\rm HM}^{\rm IIB}$ is resolved by the contributions of Euclidean D1-instantons. 

In a similar spirit, one may use \eqref{chicorr} to analyse the effect of D-instantons when approaching the infinite distance limit $\sigma \raw \infty$ in \eqref{limit}. Again, the effects of Euclidean D3, D5 and NS5-branes are negligible, so the potential \eqref{contact} should determine the metric of this region of ${\cal M}_{\rm HM}^{\rm IIB}(X)$. Notice that, even if this is a region of strong coupling, by construction the $SL(2,\mathbb{Z})$-invariant expression for \eqref{contact} provides exact results on $g_s$ \cite{RoblesLlana:2006is}.\footnote{Strictly speaking, it is the metric derived from \eqref{contact} that is $SL(2,\mathbb{Z})$-invariant, while \eqref{contact} undergoes K\"ahler transformations under the action of $SL(2,\mathbb{Z})$.} Alternatively, since the scaling \eqref{limit} takes us to the strong coupling and large Einstein frame volumes, one may consider computing the metric of this region in F-theory or, via the c-map, in the M-theory context. The latter computation was carried out in \cite{Collinucci:2009nv}, where the same expression for $\chi$ was recovered. Finally, as discussed in Appendix \ref{ap:typeIIA} one may perform the same computations as below in the type IIA mirror trajectory within ${\cal M}_{\rm HM}^{\rm IIA}(Y)$, where no strong coupling limit is involved. 

Before taking the limit, it is useful to perform a Poisson resummation on the integer $n$ in \eqref{chicorr}, that makes manifest the origin of each of the corrections. One finds   \cite{RoblesLlana:2007ae}
\begin{equation}
\chi_{\rm corr} =  \chi_{\rm pert} + \chi_{\rm WS} + \chi_{D}  \, ,
\end{equation}
where
\begin{eqnarray}
\label{chipert}
\chi_{\rm pert} & =& - \frac{\chi_E(X)}{8(2\pi)^3} \left[\zeta(3) \tau_2^2 + \frac{\pi^2}{3} \right] \, ,\\
\label{chiWS}
\chi_{\rm WS} & =& \frac{\tau_2^2}{4(2\pi)^3}\sum_{\bf{k} > 0} n_{\bf{k}}^{(0)}\text{Re}\left[\text{Li}_3\left(e^{2\pi i k_a z^a}\right)+2\pi k_a t^a \text{Li}_2\left(e^{2\pi i k_a z^a}\right)\right]\, ,\\
\label{chiD}
\chi_{\rm D} & = & \frac{\tau_2}{8\pi^2} \sum_{\bf{k}_\Lambda\neq 0}
n_{\bf{k}}^{(0)}  \sum_{m=1}^\infty \frac{|k_\Lambda z^\Lambda|}{m} \, {\rm cos}\, (2\pi m k_\Lambda \zeta^\Lambda)\, K_1 \left(2\pi m |k_\Lambda z^\Lambda|\tau_2\right) \, .
\end{eqnarray}
Here $\chi_{\rm pert}$ can be interpreted as the one-loop corrections in $\alpha'$ and $g_s$, respectively, while $\chi_{\rm WS}$ are the world-sheet instanton corrections. In the latter,  the vector $\bf{k}>0$ runs over effective homology classes $H_2^+(X,\mathbb{Z})$, and it is such that at least one of its entries $k_a \in \IN$ is non-vanishing, while  ${\rm Li}_s(x) = \sum_{r=1}^\infty r^{-s} x^r$ is the polylogarithm function. Finally, $\chi_{\rm D}$ can be interpreted as the contribution from bound states of Euclidean D1 and D(-1)-branes. To describe it we define the extended vectors $z^\Lambda = (1, z^a)$,  $\zeta^\Lambda=(\tau_1, \tau_1b^a-c^a)$ and ${\bf k}_\Lambda\ = (k_0, {\bf k}) \neq {\bf 0}$, where now $\bf{k}\geq 0$ and  $\bf{k}_\Lambda\neq 0$. Each term of the sum corresponds to a bound state of $m$ D1-branes wrapping rational curves on the class $k_a\gamma^a$ and $mk_0$ D(-1)-branes, with classical action
\begin{equation}
S_{m,\bf{k}_\Lambda} = 2\pi m  |k_\Lambda z^\Lambda|\tau_2 + 2\pi i m k_\Lambda \zeta^\Lambda\, .
\label{Smk}
\end{equation}
Clearly, most of the terms in $\chi_{\rm pert} + \chi_{\rm WS}$ quickly vanish along the trajectory \eqref{limit}, with the exception of one constant term that can also be neglected when $\chi_{\rm cl}$ is much larger. Regarding $\chi_{\rm D}$ the different terms in the sum will be negligible or not, depending on the argument of the modified Bessel function $K_1$. On the one hand, for  $x \gg 1$ we can approximate $K_1(x) \sim \sqrt{\frac{\pi}{2x}}\, e^{-x}$ and such terms can be neglected. On the other hand, for $x \ll 1$ the leading behaviour is $K_1(x) \sim \frac{1}{x}$ and therefore such terms cannot be neglected. As the argument of $K_1$ is the real part of \eqref{Smk}, this translates into the fact that instantons with large $\re\, S_{m,\bf{k}_\Lambda}$ can be neglected in $\chi_{\rm D}$, while those with small $\re\, S_{m,\bf{k}_\Lambda}$ may give a substantial correction to $\chi_{\rm cl}$. To sum up, we find that in the large volume and strong coupling limit, and more precisely as we proceed along the infinite distance trajectory \eqref{limit}, the non-negligible part of the contact potential will read
\begin{equation}
\chi = \frac{1}{12} \tau_2^2  {\cal K}_{abc}t^at^bt^c + \frac{\tau_2}{8\pi^2} \sum_{{\bf k_\Lambda}} n_{\bf{k}}^{(0)}  \sum_{m}\frac{|k_\Lambda z^\Lambda|}{m} \, {\rm cos}\, (2\pi m k_\Lambda \zeta^\Lambda)\, K_1 \left(2\pi m |k_\Lambda z^\Lambda|\tau_2\right) \, ,\label{chinonnegligible}
\end{equation}
where ${\bf k_\Lambda}= (k_0, {\bf k}) \in \IZ \times \IN^{h^{1,1}+1}\backslash 0$ and $m \in \IN$ are such that 
\begin{equation}
2\pi m |k_0 + k_a z^a|\tau_2 \ll 1\, ,
\label{cutoff}
\end{equation}
since these are the terms that will dominate the sum. Applying the above asymptotics one then obtains
\begin{equation}
\chi \sim \frac{1}{12} \tau_2^2  {\cal K}_{abc}t^at^bt^c + \frac{1}{16\pi^3} \sum_{{\bf k_\Lambda}} n_{\bf{k}}^{(0)}  \sum_{m}\frac{1}{m^2} \, {\rm cos}\, (2\pi m k_\Lambda \zeta^\Lambda)\, ,
\label{approxchi}
\end{equation}
where $m$, ${\bf k_\Lambda}$ are still such that \eqref{cutoff} holds. Of course, the set of instantons satisfying this condition depends on the value of $\sigma$ along the trajectory. Switching off the vev of the periodic fields, in particular those of $b^a$, we have that \eqref{cutoff} translates into the condition $2\pi m |k_0 e^{-\frac{3}{2} \sigma} + k_at^a (x_0) e^{-\frac{1}{2} \sigma}|\tau_{2}(x_0) \ll 1$. There will be then several towers of instantons contributing to $\chi$, with the following spectrum of actions:
\begin{table}[h]
\begin{center}
\begin{tabular}{lll}
${\bf k_\Lambda} = (k_0, \bf{0})$ & $\raw$ & $ S_{\bf k_\Lambda} \sim \frac{mk_0}{V(t)^{1/2}}$ \, ,\\
${\bf k_\Lambda} = (0, \bf{k})$ & $\raw$ & $ S_{\bf k_\Lambda} \sim \frac{m}{V(t)^{1/6}}$\, ,\\
${\bf k_\Lambda} = (k_0, \bf{k})$ & $\raw$ & $ S_{\bf k_\Lambda} \sim \frac{mk_0+ m V(t)^{1/3} }{V(t)^{1/2}}$\, .
\end{tabular}
\end{center}
\end{table}

This spectrum reproduces the mass spectrum of D-particles found in \cite{Grimm:2018ohb} for their case $d=3$, as expected from applying the c-map and mirror symmetry. Notice that in our case we have a natural way of selecting a subset of instantons within the whole tower. As dictated from the general expression for $\chi$, only those instantons satisfying \eqref{cutoff} will significantly correct the metric in ${\cal M}_{\rm HM}$. In the D-particle setup of \cite{Grimm:2018ohb}, a similar cut-off was invoked in terms of the species bound, leading to specific corrections for gauge couplings and field space metrics which then led to the Emergence Proposal. In the next section we will discuss how to address the same questions from the instanton viewpoint. 

\section{Towers of instantons and emergence}
\label{s:towers}

In the following we will explore the consequences of the instanton corrections to the hypermultiplet moduli space metric, as codified in the corrected potential \eqref{chinonnegligible}. First, the instanton effects correct the relation between the string scale and the 4d Planck mass, such that it overcomes the classical relation as we approach the infinite distance point. As a consequence, if the string scale is kept constant along the trajectory the 4d Planck mass blows up. Second, the corrections to the metric are such that they also overcome the classical metric, and render the trajectory of finite length. Finally, we analyse the corrected axion decay constants from the viewpoint of the Weak Gravity Conjecture, and show that their value is in agreement with the appearance of tensionless strings. 

\subsection{The 4d Planck mass}
\label{ss:4dplanck}

One interesting observations (see e.g. \cite{Alexandrov:2008gh}) is that the tensor potential in \eqref{contact} can be interpreted as $\chi = e^{-2\phi_4}$, where $\phi_4$ is the T-duality invariant quantity known as the four-dimensional dilaton. In particular, $\chi_{\rm cl}$ can be interpreted as the classical value for $e^{-2\phi_4}$ and $\chi$ as the quantum corrected version of the same quantity. Notice that at the classical level we have the relation 
\begin{equation}
M_{\rm P}^2 = 4\pi \chi M_s^2\,  ,
\label{Mpchi}
\end{equation}
which one may promote to the full $\chi$, giving a quantum corrected version of the relation between the Planck and string scales. Indeed, in \cite{Antoniadis:1997eg} the term ${\bf k} =0$ in \eqref{chicorr}, was argued to correct such a relation, so \eqref{Mpchi} can be seen as an extension to the full set of instantons. 

Let us see the consequences of the relation \eqref{Mpchi} in our setup, namely as we approach the infinite distance point along \eqref{limit}. Taking the approximation \eqref{approxchi} and for simplicity switching off the vevs of the periodic fields $\tau_1$, $b^a$, $c^a$, so that $\zeta^\Lambda=0$ one obtains
\begin{eqnarray}\nonumber
\chi & \sim &\frac{1}{12} \tau_2^2  {\cal K}_{abc}t^at^bt^c + \frac{1}{16\pi^3} \sum_{{\bf k_\Lambda}} n_{\bf{k}}^{(0)}  \sum_{m}\frac{1}{m^2} \\ 
& \sim & \frac{1}{12} \tau_2^2  {\cal K}_{abc}t^at^bt^c + \frac{1}{96\pi} \left[-\chi_E(X) + \sum_{{\bf k} > 0} n_{\bf{k}}^{(0)}\right]  \sum_{k_0} 1
\label{limitchi}
\end{eqnarray}
where the condition  $2\pi m |k_0 e^{-\frac{3}{2} \sigma} + k_a e^{-\frac{1}{2} \sigma}| \ll 1$ must be imposed. In the second line we have used that $\sum_{m=1}^\infty \frac{1}{m^2}= \frac{\pi^2}{6}$ converges very quickly for the first few terms, and so in practice the restriction on the sum over $m$ can be neglected. One can also see that, for a given value of ${\bf k}$, the number of  $k_0$ that satisfy the above condition is essentially always the same, and that it is similar to the $k_0$'s satisfying the condition $2\pi |k_0 | \ll e^{\frac{3}{2} \sigma}$.  As a result the sum over the D(-1) instanton number $k_0$ factors out. Finally, for large values of $\sigma$, one may replace the condition $2\pi |k_0 | \ll e^{\frac{3}{2} \sigma}$  by $|k_0| \leq e^{\left(\frac{3}{2}-\epsilon\right) \sigma}$,  with $\epsilon > 0$ parametrising the growth of the number of D(-1) instantons contributing significantly to $\chi$. Taking all this into account, we end up having an asymptotic behaviour of the form
\begin{equation}
\chi \stackrel{\sigma \raw \infty}{\sim} \frac{1}{2} \tau_2^2 V(t) + \frac{\Xi_X(\sigma)}{96\pi} \, e^{\left(\frac{3}{2}-\epsilon\right)\sigma}
\label{chilimit}
\end{equation}
where $\Xi_X = \sum_{{\bf k}>0} n_{\bf{k}}^{(0)} - \chi_E(X)$ is a growing function of $\sigma$, as the sum includes those vectors ${\bf k} > 0$ such that $2\pi k_a  \ll e^{\frac{1}{2} \sigma}$.  Again, one may replace this condition by $k_a \leq e^{\left(\frac{1}{2}-\epsilon\right) \sigma}$,  with the same value for $\epsilon$ as above. Because the actual dependence of $\Xi_X$ is related to the distribution of non-vanishing genus-zero Gopakumar-Vafa invariants $n_{\bf{k}}^{(0)}$, one may parametrise its total growth as $\Xi_X(\sigma) \sim e^{h^{1,1} \left(\frac{1}{2}-\eta\right)\sigma}$. For instance, if the $n_{\bf{k}}^{(0)}$ were non-zero for any value of {\bf k} and of similar magnitude we would have $\eta = \epsilon$, while if they were bounded one should have $\eta = 1/2$ for some $\sigma$ large enough. Plugging this result into \eqref{Mpchi},  we see that as we approach the infinite distance point the 4d Planck mass behaves as
\begin{equation}
M_{\rm P}^2 = M_{\rm P, cl}^2 + M_{\rm P, corr}^2(\sigma) = \left(2\pi \tau_2^2 V(t) +  \frac{N_{\rm sp}(\sigma)}{24} \right) M_s^2
\label{Mpcorr}
\end{equation}
where essentially $N_{\rm sp}$ is the number of instantons contributing significantly to $\chi$, and therefore to the redefinition of the 4d Planck mass. As $N_{\rm sp}$ grows rapidly with $\sigma$, it soon dominates the contribution to $M_P$, which quickly grows to infinity along the trajectory. 

Needless to say, the behaviour $M_{\rm p} \raw \infty$ goes against the initial motivation to take the specific trajectory \eqref{limit} and, in general, against the philosophy of the Swampland Program. The most natural way to circumvent this problem and have a finite 4d Planck mass for each value of $\sigma$ is to redefine the value of the string scale accordingly. In the region in which $ M_{\rm P, corr} \gg M_{\rm P, cl}$ this amounts to impose that
\begin{equation}
M_s \sim \frac{M_{\rm P}}{\sqrt{N_{\rm sp}}}
\label{sspecies}
\end{equation}
where $M_{\rm P}$ is fixed and $M_s$ and $N_{\rm sp}$ depend on $\sigma$.  If on the contrary one insists to keep the string scale fixed,  \eqref{sspecies} provides a lower bound for the 4d Planck mass in terms of $N_{\rm sp}$.

Remarkably, the rhs of this expression is similar to the one defining the species scale $\Lambda_s$ at which, in the presence of a large number of particle states $N_{\rm sp}$, gravity becomes strongly coupled \cite{Dvali:2007hz}. In our setup the number of particles is essentially replaced by the number of instantons that contribute to the divergence of the corrected Planck mass, while the cut-off scale $\Lambda < \Lambda_s$ at which the effective field theory must break down is identified with the string scale. In general, whenever the relation \eqref{Mpcorr} is valid with $N_{\rm sp}$ a rapidly growing function along a trajectory in field space, the above reasoning will apply. As $N_{\rm sp}$ grows, either one decreases the string scale or the 4d Planck mass grows like its square root.

\subsection{Removing the infinite distance}
\label{ss:infinimetric}

At the classical level the metric of ${\cal M}_{\rm HM}^{\rm IIB}$ restricted to the coordinates $(\tau_2, z^a)$ reads
\begin{equation}
\frac{1}{2} (d\phi_4)^2 + g_{a\bar{b}}dz^ad\bar{z}^{b} \, ,
\label{FSsimp}
\end{equation}
where $\phi_4$ is the classical four-dimensional dilaton and the metric along the K\"ahler coordinates is computed via a K\"ahler potential
\begin{equation}
g_{a\bar{b}} = \partial_{z^a}\partial_{\bar{z}^{b}} K \quad \quad {\rm with} \quad \quad K = - {\rm log}\, \chi_{\rm cl} \, .
\label{qcmetric}
\end{equation}
We then obtain
\begin{equation}
g_{a\bar{b}}^{\rm cl} = \frac{3}{2{\cal K}^2}\left(\frac{3}{2}{\cal K}_a{\cal K}_b -{\cal K} {\cal K}_{ab} \right)\, ,
\label{gcl}
\end{equation}
where ${\cal K} = {\cal K}_{abc}t^at^bt^c$,  ${\cal K}_a = {\cal K}_{abc}t^bt^c$, ${\cal K}_{ab} = {\cal K}_{abc}t^c$. By construction, the trajectory \eqref{limit} leaves invariant the classical 4d dilaton, and so the first factor of \eqref{FSsimp} will not contribute to the path length. Such length is computed by taking the tangent vector along the trajectory, namely 
\begin{equation}
\partial_\sigma (\tau_2, z^{a}) = \left(-\frac{3}{2} \tau_2(\sigma), i t^a(\sigma)\right)\, ,
\end{equation}
and plugging it into the second factor of \eqref{FSsimp}, obtaining a constant norm  $\| \partial_\sigma z\|^2 = \frac{3}{4}$, which is a requirement for a geodesic path.  Integrating $\| \partial_\sigma z\|$  over the range $(0, \infty)$ is equivalent to compute the proper classical distance of the corresponding direction in field space, which is obviously infinite.

However, as we proceed along the trajectory, the effect of D1/D(-1)-instantons will become more and more relevant, and will significantly modify the classical metric. As discussed in Appendices \ref{ap:metric} and \ref{ap:typeIIA}, one may capture the asymptotic behaviour of this metric by simply replacing $\chi^{\rm cl} \raw \chi$ in the above Ansatz, which implies taking $K = -2\phi_4 = {\rm log} \chi$. Then one obtains that, asymptotically
\begin{equation}
(d{\rm log} \chi)^2 \sim  \exp\left[-2\epsilon\sigma\right] (d\sigma)^2 \,,  \quad \quad g_{z \bar z}\sim \exp\left[-\left(2+\epsilon\right)\sigma\right]\,.
\label{gcorr}
\end{equation}
Notice that the asymptotic behaviour of the quantum corrected metric is the same as for the classical metric \eqref{gcl}, up to the correction given by $\epsilon$. Recall that, by the discussion of the previous subsection, this correction is by construction always non-vanishing, or else we would not be capturing the set of instantons that become relevant as we proceed along the trajectory. More precisely, for consistency it must be that this parameter lies in the range $0< \epsilon \leq 1/2$, see also the discussion in Appendix \ref{ap:metric}.

In the corrected metric, the trajectory \eqref{limit} is not strictly speaking a geodesic, because the norm of the velocity vector is no longer constant. Instead we have that
\begin{equation}
\| \partial_\sigma z\|^2 \sim e^{-\epsilon\sigma} \, ,
\end{equation}
and that this contribution dominates the vector length for sufficiently large values of $\sigma$. In this region, the parametrisation that gives a velocity vector of constant norm is instead
\begin{equation}
t^a (\rho) = \rho^{-2/\epsilon} t^a (x_0)\, , \qquad \qquad \tau_2(\rho) = \rho^{3/\epsilon} \tau_2(x_0)\, ,
\label{newlimit}
\end{equation}
with $\rho \in [0,1]$. As the domain of the new parameter is now bounded, the decompactification limit -- which now corresponds to $\rho = 0$ -- no longer is at infinite distance in field space. In terms of $\rho$ the K\"ahler coordinate metric near this point reads
\begin{equation}
g_{z \bar z}\sim \rho^{2(2+\epsilon)/\epsilon} \, .
\label{gcorrho}
\end{equation}

Finally, from the results of Appendix \ref{ap:metric} one may extract the kinetic terms for the periodic fields $\tau_1$ and $c^a$ which, in the regimes where D1/D(-1)-instanton effects are irrelevant, can be considered as axions of the compactification. The decay constant of these would-be axions evaluated at zero vev is given by 
\begin{eqnarray}
\label{gaxion00}
f_{00} & = & \rho^{(\epsilon-3)/\epsilon} M_{\rm P}\, , \\
f_{aa} & = & \rho^{(\epsilon-1)/\epsilon} M_{\rm P}\, ,
\label{gaxionaa}
\end{eqnarray}
which become infinite at $\rho=0$. Notice however that the quantum corrected metric heavily depends on $\tau_1$, $b^a$ and $c^a$ through the factor ${\rm cos}\, (2\pi m k_\Lambda \zeta^\Lambda)$, see eq.\eqref{chinonnegligible} and the expressions in Appendices \ref{ap:metric} and \ref{ap:typeIIA}. For this reason, extending the classical metric Ansatz \eqref{FSmetric} along these directions is not justified, and therefore it is not clear that one can interpret such periodic directions have infinite radius at $\rho=0$. In any case, they are clearly  not isometries of ${\cal M}_{\rm HM}^{\rm IIB}$ that can be interpreted as global symmetries in the effective theory. Despite this, an amusing fact is that the conditions $|k_0| \leq e^{\left(\frac{3}{2}-\epsilon\right) \sigma}$ and $k_a \leq e^{\left(\frac{1}{2}-\epsilon\right) \sigma}$ used above to define $\epsilon$ can be rewritten as
\begin{equation}
|k_0| \leq  \frac{f_{00}}{M_{\rm P}}\, ,\qquad \qquad k_a \leq  \frac{f_{aa}}{M_{\rm P}}\, .
\label{eWGCa}
\end{equation}
This is reminiscent of the electric Weak Gravity Conjecture for axions $f S \leq M_{\rm P}$ \cite{ArkaniHamed:2006dz}, which can be rewritten as $S^{-1} \geq f/M_{\rm P}$. Indeed, recall that \eqref{eWGCa} selects those $k_0$ and $k_a$ that corresponds to instantons whose action is small enough to have a significant contribution to $\chi$. Because increasing $(k_0, k_a)$ lowers $S^{-1}$, eq.\eqref{eWGCa} can be interpreted as a lower bound for $S^{-1}$ in terms of $f/M_{\rm P}$, much in the spirit of the WGC. Notice that, remarkably, this statement is made in terms of quantum corrected quantities $f$, $M_{\rm P}$ and not in terms their classical counterparts. For the latter, one could have already made this observation directly by inspecting \eqref{Smk}. 

While $\tau_1$ and $c^a$ can only be considered periodic coordinates in the quantum corrected metric,  the scalars $c^0$ and $d^a$ are still axions, since the instantons that would break their shift symmetry are negligible in the limit $\rho=0$. The corresponding decay constants are
\begin{eqnarray}
\tilde{f}_{00} & = & \rho^{\frac{h^{1,1}+6}{\epsilon} - 2 h^{1,1}- 3} \, M_{\rm P}\, , \\
\tilde{f}_{aa} & = & \rho^{\frac{h^{1,1}+4}{\epsilon} - 2h^{1,1}-3} \, M_{\rm P}\, ,
\end{eqnarray}
which vanish as $\rho \raw 0$. As these are isometries of ${\cal M}_{\rm HM}^{\rm IIB}$, they a priori represent global symmetries in the 4d effective field theory, in conflict with standard wisdom \cite{Banks:1988yz,Banks:2010zn}. However, in the same limit where they become exact symmetries, 4d strings become tensionless, as we now turn to discuss.

\subsection{Tensionless strings}
\label{ss:strings}

Just like when proceeding along the trajectory \eqref{limit} several towers of D1/D(-1)-instantons dramatically decrease their action, the same is true for the tension of certain D-strings. In particular, bound states of D3/D1-branes wrapping holomorphic cycles are seen as 4d strings which, in the limit $\sigma \raw \infty$, become tensionless.\footnote{For a more general analysis of the spectrum 4d strings in different limits of infinite distance see \cite{tensionless}. } It is in fact instructing to analyse the behaviour of their (classical) tension in the quantum corrected trajectory \eqref{newlimit}. The tensions of a D1-brane pointlike in $X$ and a D3-brane wrapping a holomorphic 2-cycle of $X$ is given by
\begin{eqnarray}
\label{TD1}
T_{D1} & = & \rho^{3/\epsilon} M_s^2 = \rho^{\frac{h^{1,1}+6}{\epsilon} -2 (h^{1,1}+ 1)} \, M_{\rm P}^2\, , \\
T_{D3} & = & \rho^{1/\epsilon} M_s^2 = \rho^{\frac{h^{1,1}+4}{\epsilon} -2 (h^{1,1}+ 1)} \, M_{\rm P}^2 \, ,
\label{TD3}
\end{eqnarray}
where we have used eq.\eqref{chiscaling} to relate the string scale with the corrected Planck scale. In both cases, for $0< \epsilon \leq 1/2$ the exponent of $\rho$ is positive, and so these tensions decrease to zero as we reach the limit $\rho \raw 0$. 

Let us now compare these tensions with the decay constants of the axions which are magnetically charged under each of these strings. One finds that
\begin{eqnarray}
\label{TD1f}
T_{D1} & = & \tilde{f}_{00} M_{\rm P} \, \rho \,  \leq \, \tilde{f}_{00} M_{\rm P} \, ,  \\
T_{D3} & = & \tilde{f}_{aa} M_{\rm P} \, \rho \,  \leq \, \tilde{f}_{aa} M_{\rm P}\, , 
\label{TD3f}
\end{eqnarray}
in agreement with the magnetic version of the Weak Gravity Conjecture for axions \cite{Reece:2018zvv,Hebecker:2017uix} (see also \cite{Hebecker:2017wsu}), with an intriguing extra factor of $\rho$. In this sense, one may interpret $\sqrt{T_{D1}}$ as the cut-off scale of the effective field theory, lying below $\sqrt{T_{D3}}$ and $M_s$. As this cut-off approaches zero at the same point where the field space metric becomes singular, it is tempting to speculate whether the tensionless string states may have a similar role to the massless hypermultiplet in the conifold case, and if they could also shed light into the asymptotics of the axion decay constants \eqref{gaxion00} and \eqref{gaxionaa} which, despite instanton effects, still become infinite at $\rho =0$. 

One important question in this respect is the precise spectrum of tensionless strings and the corresponding excitations that may affect the hypermultiplet metric. First, the complete set of stable bound states is not obvious to determine, even if one resorts to monodromy arguments \cite{Grimm:2018ohb,tensionless}. Second, in this regime the tension of the strings could be modified by the same non-perturbative effects that removed the infinite distance, and this could in turn prevent some D3/D1-strings to become tensionless. Because of the magnetic version of the WGC for axions, one would expect at least some D-strings to become tensionless, given that the rhs of \eqref{TD1f} and \eqref{TD3f} are computed for quantum corrected quantities. Finally, one should understand which subset of massive states should be integrated out to connect with physics at the IR scale.

In any event, it would be interesting to see if the effect of tensionless strings can be incorporated to that of instantons, perhaps via some dual description of our setup. This could involve performing an $SL(2,\mathbb{Z})$ transformation to a weakly coupled regime, in which the tensionless D1-strings become fundamental strings. Alternatively, one could consider a chain of dualities that maps the tensionless D3-strings into tensionless fundamental strings, much in the spirit of \cite{Lee:2018urn,Lee:2018spm,Lee:2019tst}. Indeed, notice that the appearance of tensionless D3-strings in our setup is reminiscent of the constructions in \cite{Lee:2018urn,Lee:2018spm,Lee:2019tst} upon replacing the vanishing gauge coupling constants by infinite axion decay constants. Therefore, it would be interesting to see the result of combining both pictures, a problem to which we hope to return in the future.

\subsection{D-particles and emergence}
\label{ss:Dpart}

At the classical level, the type IIB hypermultiplet moduli space ${\cal M}_{\rm HM}^{\rm IIB}(X)$  metric reads \cite{Ferrara:1989ik}
\begin{eqn}
ds^2_{{\rm HM, IIB}} = \frac{1}{2} d\phi_4 + ds^2_{{\rm VM, IIA}} + ds^2_{{\rm axions}}\, ,
\label{HMmetric}
\end{eqn}
where $\phi_4$ is the four-dimensional dilaton, $ds^2_{{\rm VM, IIA}}$ depends on the complex coordinates $z^a$ and is identical to the type IIA vector multiplet metric on the same Calabi-Yau $X$, and $ds^2_{{\rm axions}}$ is the metric along the axionic directions $b^0$, $c^0$, $c^a$, $d^a$, see \eqref{FSmetric} for a more detailed expression. 

The D-instanton action \eqref{Smk} expressed in these coordinates is (see e.g. \cite{Pioline:2009ia}) 
\begin{eqn}
S_{m,\bf{k}_\Lambda} = \frac{2\pi m}{g_4} e^{{\cal K}/2} | Z_{\bf{k}_\Lambda} | + 2\pi i m k_\Lambda \zeta^\Lambda\, ,
\label{Smk2}
\end{eqn}
where $g_4 = e^{\phi_4}$. Here $e^{{\cal K}/2} | Z_{\bf{k}_\Lambda} |$ is the normalised central charge function that depends on the complexified K\"ahler moduli $z^a$ and that, when we compactify type IIA on $X$, governs the mass of the D-particles, see \cite{Corvilain:2018lgw}. Clasically, we thus find that the magnitude of the instanton correction $|e^{-S_{m,\bf{k}_\Lambda}}|$ depends on the coordinates corresponding to the first two factors in \eqref{HMmetric}. We may thus conceive two types of infinite geodesics, under which the D-instanton corrections behave quite differently:

\begin{itemize}

\item[-] $\phi_4 \raw \infty$, $z^a$ constant: \ All D-instanton actions vanish asymptotically $S_{m,\bf{k}_\Lambda} \raw 0$, and the same is true for the NS5-brane instanton. In the large volume region one may also neglect $\alpha'$ and $g_s$ corrections and describe the trajectory classically. We thus find an infinite trajectory in which $M_s/M_{\rm P} \simeq  M_s/M_{\rm P, cl}  = e^{-\phi_4}$ tends to zero exponentially. The SDC is satisfied by the tower of string states. 

\item[-] $\phi_4$ constant, $z^a$ approach infinite distances in $ds^2_{{\rm VM, IIA}}$: \ Then $\re \, S_{m,\bf{k}_\Lambda}$ behaves like the spectrum of D-particles when we approach infinite distance points in $ds^2_{{\rm VM, IIA}}$, except that now we have the additional integer $m$ indexing the instanton expansion. Therefore, for each tower of D-particles that becomes massless exponentially fast along an infinite trajectory in $ds^2_{{\rm VM, IIA}}$ we will have a tower of towers of instantons whose action will have the same behaviour. We can therefore borrow the results in \cite{Grimm:2018ohb,Grimm:2018cpv,Corvilain:2018lgw} to conclude that at infinite distance points of this sort the D-instanton corrections may substantially modify the moduli space metric. Finally, if $\phi_4$ is fixed at a large enough value, the NS5-instanton effects will be negligible. 

\end{itemize}

The trajectory \eqref{limit} can be seen as a particular case of these (classical) infinite distance trajectories, in which D1/D(-1)-instantons effects become dominant over all other corrections to the classical metric. It seems reasonable to expect that, in all of these cases, the D-instanton quantum corrections to 4d dilaton dominate over the others, and one obtains a relation of the form \eqref{sspecies}, with $N_{\rm sp}$ growing exponentially fast along the trajectory. Then, if the distance remains of infinite length after quantum corrections, the SDC would be satisfied by the tower of fundamental string states. In fact, quite probably there will be further 4d strings whose tension in 4d Planck units decay even faster, as it happens in eqs.\eqref{TD1f}, \eqref{TD3f} in our example. Then there will be several towers of string excitations satisfying the Swampland Distance Conjecture. 

On the contrary it could occur that the same result found for \eqref{limit} applies to other trajectories, namely that the instanton effects render finite the classically infinite length. If this was true for all the infinite distance trajectories of this kind then, naively, in the quantum corrected hypermultiplet moduli space only the direction $\phi_4 \raw \infty$ would remain of infinite distance. It would be very interesting to gather further evidence on whether this may actually be the case.

One interesting direction to gain some intuition on this matter would be to apply the c-map to the above setup, in order to interpret it in terms of D-particles. In order to do so, one must further compactify the theory on a circle, and apply T-duality along such new compact direction. One then obtains type IIA compactified on $X \times S^1$, where the $S^1$ has the dual radius. The moduli space \eqref{HMmetric} arises from compactifying the type IIA Vector Multiplet sector on such $S^1$. The axions arise from the Wilson lines of the 4d gauge bosons along the circle, as well as from dualising the 3d gauge bosons. The direction $\phi_4$ corresponds to the radius of the circle. In fact by T-duality one obtains the relation \cite{Seiberg:1996ns}
\begin{eqn}
g_4^{\rm IIB} \,  = \,  \frac{g_4^{\rm IIA}}{2 \pi M_s R_{\rm IIA}}\, ,
\end{eqn}
which helps to identify type IIB D-instantons with type IIA D-particles. Indeed, one may take \eqref{Smk2} and translate it to type IIA vector multiplet quantities as
\be
\re\, S_{m,\bf{k}_\Lambda}^{\rm IIB} \, = \, 2\pi m R_{\rm IIA} m_{\bf{k}_\Lambda}^{\rm IIA}
\ee
where $m_{\bf{k}_\Lambda}^{\rm IIA}$ is the mass of the type IIA D-particle wrapping the internal cycle corresponding to ${k}_\Lambda$ in units of type IIA 4d Planck mass. As in \cite{Seiberg:1996ns}, one then identifies the IIB D-instanton in the sector $m,\bf{k}_\Lambda$ with the IIA D-particle from the sector ${\bf{k}_\Lambda}$ whose worldline wraps $m$ times the $S^1$. Interestingly, the condition that makes the IIB instantons correct significantly the metric, namely \eqref{cutoff} or more generally $S_{m,\bf{k}_\Lambda}^{\rm IIB} \ll 1$, translate in the type IIA side to
\be
m_{\bf{k}_\Lambda}^{\rm IIA} \ll 1/R_{\rm IIA}\, ,
\ee
so these are D-particles way below the 3d $\raw$ 4d Kaluza-Klein scale, and may be seen as a tower of particles in 3d. Following again \cite{Seiberg:1996ns}, one concludes that one should be able to generate the quantum corrected 4d hypermultiplet metric by a one-loop computation of an infinite tower of 3d D-particles, and that such integral should not receive further corrections. Needless to say, this picture is very suggestive from the viewpoint of the Emergence Proposal, as it allows to recover the IR metric of a compactification via integrating out states. Notice that in this case the said one-loop integral should be strictly thought as the running of couplings created by virtual particles, as actual ones would generate a deficit angle, and it is not clear how to accommodate an infinite tower of them. In this sense, our result could interpreted as that, unlike in 4d, towers of light particles in 3d do not necessarily generate infinite distances. In fact, it would be very interesting to show whether no infinite distances exist or emerge in the IR moduli space of 3d quantum gravitational theories. If such infinite distances were absent in both cases, it would not only constitute further evidence for the Emergence Proposal, but it would also provide an interesting extension of the SDC for the three-dimensional case. Indeed, one would then conclude that the SDC is trivially satisfied in 3d Minkowski. There would be no infinite distances because there is no infinite tower of particles that could become light when approaching them.

\section{Conclusions}
\label{s:conclu}

In this paper we have analysed the hypermultiplet moduli space metric of type II string CY compactifications, along a large volume, strong coupling trajectory. At the classical level, such a trajectory is of infinite distance, and corresponds to a decompactification limit in F/M-theory. At the quantum corrected level, the metric is heavily modified as we proceed along the trajectory, rendering such a decompactification limit at finite distance. The microscopic mechanism behind this effect is an infinite amount of D-instantons with rapidly decreasing action as we proceed along the trajectory. Their effect on the metric can be codified in the contact potential \eqref{approxchi}, which in addition measures the evolution of the quotient $M_s/M_{\rm P}$ along the said trajectory. We find that this quotient vanishes at the trajectory endpoint, even if it is at finite distance. If $M_{\rm P}$ is kept fixed then the string scale must be lowered as we proceed along the trajectory, and if $M_s$ is kept fixed then gravity effectively decouples at the trajectory endpoint. In hindsight, this seems a rather universal behaviour for points in moduli space where the effective theory displays a large number of light resonances and/or instantons with small action. Points of this sort in field space presumably contain important information with respect to the limits of the corresponding effective field theory, being related to the Swampland Distance Conjecture and certain aspects of the Emergence Proposal. 

As we have discussed in section \ref{ss:Dpart}, one can divide the infinite geodesics in the classical Hypermultiplet moduli space in two classes. The first one is  weak coupling limit in which the four-dimensional dilaton eventually vanishes. In this limit the D-instanton corrections to the metric are negligible, and if in addition we are in a large volume regime so will be the ones coming from worldsheet and NS5-branes instantons. One can then see that this geodesic direction remains infinite, and that the tower of states descending to zero mass exponentially fast are the tower of fundamental string states. This already implies that the hypermultiplet moduli space is non-compact, in agreement with the results in \cite{Kachru:1995wm,Aspinwall:1998bw,Louis:2011aa}. 

The other geodesics of infinite length in the classical metric of ${\cal M}_{\rm HM}^{\rm IIB/IIA}$ are directly related to those found in the vector multiplet moduli space ${\cal M}_{\rm VM}^{\rm IIA/IIB}$ of the same Calabi-Yau. By keeping the four-dimensional dilaton fixed one obtains the same dependence for the spectrum of instanton actions in  ${\cal M}_{\rm HM}^{\rm IIB/IIA}$ as for the spectrum of D-particle masses in $M_{\rm P}$ units in ${\cal M}_{\rm VM}^{\rm IIA/IIB}$, except that there is a tower of instantons per each D-particle. One can then borrow the results of \cite{Grimm:2018ohb,Grimm:2018cpv,Corvilain:2018lgw} to argue that at each infinite distance point there are towers of towers of instantons with vanishing action, and therefore significant corrections to the classical metric. In this paper we have considered a particular, universal geodesic of classical infinite distance and shown that, upon taking into account the relevant D-instanton effects, the distance is rendered finite in the quantum corrected metric. A natural extension of this analysis would be to consider other trajectories of this sort, for instance by using $SL(2,\mathbb{Z})$ duality and the Fourier-Mukai transform to access infinite distance points at weak coupling and/or small volume, as in \cite{Corvilain:2018lgw}. For these remaining trajectories it could be that the infinite distance is also removed by quantum corrections. Alternatively, it could happen that it is only removed for some of them, perhaps depending of the type of infinite distance according to the classification in \cite{Grimm:2018ohb,Grimm:2018cpv,Corvilain:2018lgw}. In either case, it seems quite reasonable that the Swampland Distance Conjecture is satisfied for each of the trajectories. Indeed, even if the infinite distance is not removed, the reasoning taking us to the relation \eqref{Mpcorr} should still apply. As this result implies a string scale decreasing exponentially fast along the trajectory, it provides a tower of string resonances satisfying the conjecture. In fact, one would expect that even a lower tower of 4d D-string resonances is present in these regions, satisfying the magnetic Weak Gravity Conjecture for axions not only in the classical metric but also in the quantum corrected one, as we checked in our example. Finally, if quantum corrections are able to remove infinite distance points, they may also change the curvature around them. Indeed, one of the conjectures of \cite{Ooguri:2006in} states that around points of infinite distance the scalar curvature must be negative, this being related with the volume of the moduli space being finite. It would be interesting to see if the removal of such points is connected to a change in the local curvature. 

This picture also provides further support for the recent Emergence Proposal \cite{Harlow:2015lma,Grimm:2018ohb,Heidenreich:2017sim,Heidenreich:2018kpg,Palti:2019pca}. Indeed, as discussed in section \ref{ss:Dpart}, for the infinite geodesics with constant four-dimensional dilaton the relevant D-instanton effects that correct the metric significantly should be captured, via the c-map, in the one-loop effects of a tower of 3d virtual particles. In fact, the most relevant terms of the whole quantum corrected metric should correspond to the IR metric generated by such one-loop corrections. 

Finally, another important direction would be to check if similar effects can occur in the context of ${\cal N} =1$ compactifications. There the general corrections to the K\"ahler metrics are less understood, but some of them may be analysed in certain constructions. In particular it would be interesting to see if towers of non-perturbative effects become relevant near infinite distance points in Calabi-Yau orientifold compactifications, whose K\"ahler metrics can be partially understood in terms of the ones analysed here. As in this context non-perturbative effects are oftentimes invoked to achieve realistic features, understanding their interplay with quantum gravity could be essential to sharpen the predictive power of the string Landscape. 

\newpage

\bigskip

\centerline{\bf \large Acknowledgments}

\bigskip

We would like to thank Sergei Alexandrov, Florent Baume, \'Alvaro Herr\'aez, Luis E. Ib\'a\~nez, Eran Palti, Raffaele Savelli, \'Angel Uranga, Timo Weigand, Irene Valenzuela and Cumrun Vafa for useful discussions. This work is supported by the Spanish Research Agency (Agencia Estatal de Investigaci\'on) through the grant IFT Centro de Excelencia Severo Ochoa SEV-2016-0597 and by the grant FPA2015-65480-P from MINECO/FEDER EU. MW received funding from the European Union's Horizon 2020 research and innovation programme under the Marie Sklodowska-Curie grant agreement No. 713673. The work of MW also received the support of a fellowship from ”la Caixa” Foundation (ID 100010434) with fellowship code 102A0122.

\appendix

\section{The c-map}
\label{ap:cmap}
In this appendix, we review the c-map construction \cite{Cecotti:1988qn,Ferrara:1989ik} used in the main text to relate the analysis of instanton corrections in the hypermultiplet moduli space of type IIB string theory to 1-loop corrections of the vector multiplet moduli space in type IIA string theory. The hypermultiplet moduli space $\mathcal{M}_\text{HM}^{\rm IIB}(X)$ of IIB string theory compactified on a CY $X$ can be described as a quaternionic K\"ahler manifold of quaternionic dimension $h^{1,1}(X)+1$. At the same time, the vector multiplet moduli space $\mathcal{M}_\text{VM}^{\rm IIA}(X)$ of IIA string theory on the same CY $X$ is a projective special K\"ahler manifold of complex dimension $h^{1,1}(X)$. The two moduli spaces $\mathcal{M}_\text{HM}^{\rm IIB}(X)$ and $\mathcal{M}_\text{VM}^{\rm IIA}(X)$ can now be related to each other using the c-map, which associates to every projective special K\"ahler manifold a quaternionic K\"ahler manifold in the following way:  

Consider the 4d $\mathcal{N}=2$ effective supergravity theory as obtained from a type IIA CY compactification and further compactify on an additional circle to get a 3d $\mathcal{N}=4$ theory. The 4d moduli space, which is a product space 
\begin{equation*}
    \mathcal{M}_{\rm 4d}=\mathcal{M}_\text{VM}^{\rm IIA}\times \mathcal{M}_\text{HM}^{\rm IIA}\,,
\end{equation*}
gets modified due to additional scalars which arise in 3d. In particular, these are one scalar~$R$ from the metric (the radius of the $S^1$) and the $h^{1,1}(X)+1$ Wilson line moduli~$\zeta_\Lambda$ of the 4d graviphoton and vector multiplets along the $S^1$. Moreover, dualising the 3d graviphoton, vector multiplet and KK vector gives rise to $h^{1,1}+2$ additional scalars $\left(\tilde \zeta^\Lambda, \sigma \right)$ in 3d. Due to the $\mathcal{N}=4$ supersymmetry in 3d, the moduli space after compactification to 3d is still of product form 
\begin{equation*}
     \mathcal{M}_3=\mathcal{M}_\text{twisted}^{\rm IIA}\times \mathcal{M}_\text{HM}^{\rm IIA}\,.
\end{equation*}
While $\mathcal{M}_\text{HM}^{\rm IIA}$ is independent of the radius of the $S^1$ and is thus the same in 3- and 4-dimensions, the scalars in the 4d vector multiplet pair with the new scalars to form twisted hypermultiplets. The latter space has real dimension $4\left(h^{1,1}+1\right)$ and is a quaternionic-K\"ahler manifold. The compactification to 3d thus gives us a map 
\begin{align}
    c: \;\begin{matrix}\mathcal{M}_\text{VM}^{\rm IIA} &\longrightarrow &\mathcal{M}_\text{twisted}^{\rm IIA} \\
    \left(z^a, A_\mu\right) &\longrightarrow &\left(z^a, \zeta_\Lambda, \tilde\zeta^\Lambda, R, \sigma \right)\end{matrix}\,,
\end{align}
where $z^a$ are the complexified K\"ahler moduli of $X$, between the 4d and 3d moduli spaces: the c-map. Recall that the metric of a projective special K\"ahler manifold can be determined  by a degree 2 homogeneous prepotential $F(x^I)$. The c-map now associates to such a prepotential a quaternionic K\"ahler metric $G_{ij}$ with the special property that the embedding of $\mathcal{M}_\text{VM}^{\rm IIA}$ inside $\mathcal{M}_\text{twisted}^{\rm IIA}$ is a totally geodesic submanifold \cite{Ferrara:1989ik}. 

Let us now perform a T-duality transformation to type IIA on $X\times S^1$ along the additional $S^1$. This gives type IIB on $X\times \tilde S^1$, with $\tilde S^1$ the T-dual circle, while exchanging the hyper- and twisted moduli spaces. Thus we can identify the IIB hypermultiplet moduli space $\mathcal{M}_\text{HM}^{\rm IIB}$ with $\mathcal{M}_\text{twisted}^{\rm IIA}$, the image of $\mathcal{M}_\text{VM}^{\rm IIA}$ under the c-map. In particular, the IIB RR axions get identified with the Wilson line moduli $\zeta_\Lambda$ and scalar duals $\tilde \zeta^\Lambda$ of the IIA graviphoton and vector multiplets. Moreover, this T-duality maps D-brane particles arising in the IIA compactification from D2-branes wrapped along holomorphic 2-cycles and with momentum along the $S^1$ to D1-brane instantons on the IIB side wrapped along the same 2-cycle. As a result, 1-loop corrections of these D-brane particles to the vector multiplet moduli space of the IIA compactification are captured in the c-dual IIB hypermultiplet moduli space by the corresponding D-brane instanton corrections.   

A non-zero vev for the axions in the IIB hypermultiplet moduli space corresponds using the c-map to non-zero Wilson lines of the graviphoton and vector multiplets in IIA. When calculating 1-loop corrections to the vector multiplet moduli space of IIA in 4D, the gauge fields are assumed to have vanishing Wilson lines in all extended directions. Therefore, the 1-loop corrections for the vector multiplet moduli space in 4D should correspond to instanton corrections in IIB for vanishing axions vevs. 

\section{The conifold limit}
\label{ap:conifold}
The intuition for resolving the infinite distance in the hypermultiplet moduli space of type IIB CY compactifications is based on the same physical principle that resolves the conifold singularity in the same moduli space \cite{Strominger:1995cz}. In this appendix we discuss the resolution of the conifold singularity of the type IIB hypermultiplet moduli space due to D1/D(-1)-brane instantons, following a slightly different approach to the one taken in \cite{Saueressig:2007dr}. 

Consider the instanton corrected expression for the K\"ahler potential of the type IIB hypermultiplet moduli space (\ref{contact}) and let us focus in this appendix on the case $h^{1,1}(X)=~1$ with a single complexified K\"ahler modulus $z$. The moduli space conifold singularity now corresponds to $z\rightarrow 0$ which we want approach in the weak-coupling limit $\tau_2\rightarrow \infty$ while keeping $\tau_2z $ finite. In this limit, the classical contribution to the K\"ahler potential, being proportional to the volume, vanishes while worldsheet and D-brane instantons become important.  

The perturbative, $\alpha'$ and worldsheet instanton corrections to the K\"ahler potential read
\begin{eqn}
    \chi_\text{WS+pert.}= &-\frac{\chi_X \zeta(3)}{8(2\pi)^3}\tau_2^2-\frac{\chi_X}{192 \pi}\\
    &+\frac{\tau_2^2}{4(2\pi)^3}\sum_{k_a\gamma^a\in H^+_2(X)} n_{k_a}^{(0)}\text{Re}\left[\text{Li}_3\left(e^{2\pi i k_a z^a}\right)+2\pi k_a t^a \text{Li}_2\left(e^{2\pi i k_a z^a}\right)\right]\,,
\end{eqn}
where $\text{Li}_k$ is the $k$'th Polylogarithm. The contribution of the worldsheet instantons to the hypermultiplet metric is then given by 
\begin{eqn}
    g_{z \bar z}^\text{WS}\equiv -\partial_z\partial_{\bar z} {\rm log} \chi_\text{WS+pert.}&=-\frac{1}{\chi_\text{WS+pert.}}\frac{\tau_2^2}{16 \pi}\sum\limits_{k_1}n_{k_1}^{(0)} k_1^2 \left[\log\left(1-e^{2\pi i k_1 z}\right)+c.c.\right]+\dots\\
    &\simeq \frac{1}{\chi_\text{WS+pert.}}\frac{\tau_2^2}{16 \pi}\sum\limits_{k_1}n_{k_1}^{(0)} k_1^2 \log\left(\frac{(2\pi)^2}{z\bar z}\right)+\dots \,,
\end{eqn}
where the $\dots$ represent subleading terms that we disregard here.

The other important contribution to the metric comes from the D1-brane instantons. As argued in appendices \ref{ap:metric} and \ref{ap:typeIIA}, in the regions of the moduli space where only D(-1) and/or D1-instantons become very relevant, one may capture their effect on the metric by means of the expression \eqref{gzzexact}. In the following we will demonstarate this observation in the simple case of the conifold, recovering the results of \cite{Saueressig:2007dr} from it.

The instantons that become important in the conifold limit wrap the $S^2$ that shrinks to zero size at the conifold point. For these D1-brane instantons in the limit $k\rightarrow 0$ the argument of the Bessel function $K_0$
\begin{equation}
    |k_\Lambda z^\Lambda|\tau_2= \left|z+k_0\right| \tau_2 \,,
\end{equation}
is clearly dominated by the instanton with D(-1) charge $k_0=0$. Hence the leading piece of the hypermultiplet moduli space metric close to the conifold point at $z\rightarrow 0$ is given by 
\begin{eqn}
      g_{z \bar z}^\text{leading}\simeq \frac{1}{\chi_\text{WS+pert.}}\frac{\tau_2^2}{4\pi} \left\{\frac{1}{4\pi} \log\left(\frac{(2\pi)^2}{z\bar z}\right)+\frac{1}{2\pi} \sum\limits_{m=1}^\infty \cos(2\pi m\zeta) K_0\left(2\pi | mz| \tau_2\right)\right\}\,,\label{Conifoldmetric}
\end{eqn}
where $\zeta$ is the axion obtained by reducing the RR 2-form over the shrinking $S^2$. This result for the metric agrees with what was found for the conifold metric in \cite{Ooguri:1996me} in the mirror dual type IIA hypermultiplet moduli space and can be re-expressed as
\begin{align}
    g_{z\bar z}=\frac{1}{4\pi} \sum\limits_{n=-\infty}^\infty\left[ \frac{1}{\sqrt{(\zeta -n)^2-\tau_2^2z \bar z )}}-\frac{1}{|n|}\right]+ \text{const.}\,, \label{OoguriVafametric}
\end{align}
which is regular as $z\rightarrow 0$\,. 

We thus observe that the logarithmic divergence of the field-space metric near the conifold point, as caused by the worldsheet instantons, gets resolved by the corrections due to D1-brane instantons.

\section{Quantum corrected metric components}
\label{ap:metric}
In this section we collect the expressions for the relevant components of the hypermultiplet moduli space metric as calculated from $\chi$, seen as a K\"ahler potential. We are mainly interested in the contributions to the metric that are non-negligible in the large volume/strong coupling limit considered in Section  \ref{s:typeIIB}. 

Classically, the metric on the entire hypermultiplet can be calculated from the K\"ahler potential $\chi_{\rm cl}$ along the K\"ahler submanifold spanned by the complexified K\"ahler coordinates $z^a$. This metric, up to order one constants, is given by \cite{Ferrara:1989ik}
\begin{eqn}
    h_{uv} dq^u dq^v\supset& \frac{\left(d\log \chi_\text{cl}\right)^2}{8}+\frac{1}{2\tau_2^2} \left(d\tau_1\right)^2+g_{a \bar b}dz^ad \bar z^b +\frac{g_{a\bar b}}{\tau_2^2}d \zeta^a d \zeta^b+\frac{\tau_2^2}{\chi_\text{cl}^2}g^{a\bar{b}}d\tilde \zeta_a d\tilde \zeta_b \\& + \frac{1}{\chi_\text{cl}^2} \left(\tau_2^2\left(d \tilde \zeta_0\right)^2+\left(d \sigma\right)^2\right) \,.\label{FSmetric}
\end{eqn}
where recall that 
\begin{eqn}
    \zeta^a =\tau_1b^a-c^a \,,\;\;\; \tilde \zeta_a &= d_a -\frac{1}{2}\mathcal{K}_{abc} b^b \zeta^c \,,\;\;\;\; \tilde \zeta_0 =c^0+\frac{1}{6}\mathcal{K}_{abc}b^a b^b \zeta^c\,\\
    \sigma&=-2\left(b^0+\frac{1}{2}c^0 \tau_1\right)-c_a\zeta^a+\frac{1}{6}\mathcal{K}_{abc}b^a c^b \zeta^c\,,
\end{eqn} 
and $\chi_{\rm cl}$ is the type IIB 4d dilaton, given by \eqref{chiclas}. The metric $g_{a\bar b}$ for the complexified K\"ahler moduli can be obtained by taking derivatives with respect to the K\"ahler potential $K= - {\rm log}\, \chi_{\rm cl}$.

We now consider the same Ansatz but, instead of using $\chi_{\rm cl}$, we apply it to the quantum corrected,  SL(2,$\mathbb{Z}$)-invariant 4d dilaton $\chi$ in  \eqref{contact}, and more precisely to the expression \eqref{chinonnegligible} that is relevant in the limit taken in section \ref{ss:geodesic}. Such a prescription to compute the instanton-corrected metric in this specific region in moduli space is motivated by the results of Appendix \ref{ap:conifold},  because it allows to recover well-known results in the literature that are based on the same physical principle. More importantly, they are justified by comparing the results below with the exact hypermultiplet metric results computed in the type IIA mirror symmetric setup, see Appendix \ref{ap:typeIIA}.

Following this approach, the metric components along the K\"ahler coordinates are
\begin{eqn}
       &g_{z^a \bar z^b}=-\partial_{z^a}\partial_{\bar z^b} \log \chi \label{gzzexact} \\
     =&\frac{1}{\chi^2}\left[-\frac{\tau_2^2}{8}i\,\mathcal{K}_{ace}t^ct^e-\frac{\tau_2^2}{8\pi} \sum_{{\bf k_\Lambda}} n_{\bf{k}}^{(0)}  \sum_{m} k_\Lambda \bar z^\Lambda k_a\cos(2\pi m k_\Lambda \zeta^\Lambda) K_0\left(2\pi m |k_\Lambda z^\Lambda| \tau_2\right) \right]\\
      &\times \left[\frac{\tau_2^2}{8}i\,\mathcal{K}_{bdf}t^dt^f-\frac{\tau_2^2}{8\pi} \sum_{{\bf k_\Lambda}} n_{\bf{k}}^{(0)}  \sum_{m} k_\Lambda z^\Lambda k_b\cos(2\pi m k_\Lambda \zeta^\Lambda) K_0\left(2\pi m |k_\Lambda z^\Lambda| \tau_2\right) \right]\\
     &-\frac{1}{\chi}\left[\frac{\tau_2^2}{8}\mathcal{K}_{abc}t^c-\frac{\tau_2^2}{8\pi} \sum_{{\bf k_\Lambda}} n_{\bf{k}}^{(0)}  \sum_{m} k_a k_b\cos(2\pi m k_\Lambda \zeta^\Lambda) K_0\left(2\pi m |k_\Lambda z^\Lambda| \tau_2\right)\right.\\
      &\left.+\frac{\tau_2^2}{16\pi}\sum_{{\bf k_\Lambda}} n_{\bf{k}}^{(0)}  \sum_{m} k_a k_b 2\pi m |k_\Lambda z^\Lambda|\tau_2 \cos(2\pi m k_\Lambda \zeta^\Lambda) K_1\left(2\pi m |k_\Lambda z^\Lambda| \tau_2\right)\right]\, ,
\end{eqn}
which to great extent specifies the metric \eqref{FSmetric}, and in particular the length of the trajectory \eqref{limit}. The other relevant contribution comes from the first factor in \eqref{FSmetric}. Recall that at the classical level the 4d dilaton $\sim \chi_\text{cl.}$ stays constant along \eqref{limit}, and so this term does not contribute to the path length. As discussed in the main text, this changes dramatically once that D1/D(-1)-instanton effects are taken into account. One may compute this contribution by rewriting the first term in \eqref{FSmetric} as
\begin{align}
    \frac{\left(d\log \chi\right)^2}{8} = \frac{1}{8\chi^2}\left(\frac{\partial \chi}{
    \partial \sigma}\right)^2 \left(d\sigma\right)^2 \,, \label{dchidsigma}
\end{align}
where $\chi$ as a function of $\sigma$ is directly obtained from  \eqref{chinonnegligible}
\begin{eqn}
   \chi(\sigma)&= \text{const.} + \frac{1}{8\pi^2} \sum_{{\bf k_\Lambda}} n_{\bf{k}}^{(0)}  \sum_{m}\frac{f(\sigma)}{m} \, {\rm cos}\, (2\pi m k_\Lambda \zeta^\Lambda)\, K_1 \left(2\pi m f(\sigma)\right)\,,\\
    \text{with}\;\;\;&f(\sigma)= \tau_2(x_0) e^{-\frac{3}{2}\sigma}\left|k_a\left(b^a + it^a(x_0) e^{\sigma}\right)\right| \,.
\end{eqn}
Here we have introduced the function $f(\sigma)$ that captures the entire $\sigma$ dependence of $\chi$. We can now evaluate $\partial_\sigma \chi$ and obtain
\begin{eqn}
    \partial_\sigma \chi&=-\frac{1}{4\pi} \sum_{{\bf k_\Lambda}} n_{\bf{k}}^{(0)}  \sum_{m}f(\sigma) \partial_\sigma f(\sigma)\, {\rm cos}\, (2\pi m k_\Lambda \zeta^\Lambda) K_0 \left(2\pi m f(\sigma)\right)\,,\\
    f(\sigma)\partial_\sigma f(\sigma)&=-\frac{3}{2} f^2(\sigma)+\tau_2^2(x_0)e^{-3\sigma} \left(k_a t^a(x_0)e^\sigma\right)^2\, , 
\end{eqn}
which can now be inserted into \eqref{dchidsigma} to obtain the said contribution to the path length.

To analyse the asymptotic behaviour of the metric components \eqref{gzzexact} and \eqref{dchidsigma} in the limit \eqref{limit}, let us first rewrite the sums over the Bessel functions $K_0$ and $K_1$ by performing a Poisson resummation over the integer $m$. We therefore use 
\begin{align}
    \sqrt{\alpha}\left(\frac{1}{2}f(0)+\sum\limits_{n=1}^{\infty}f(\alpha n)\right)= \sqrt{\beta }\left(\frac{1}{2}F_c(0)+\sum\limits_{n=1}^{\infty}F_c(\beta n)\right)\,, \;\;\; \alpha\beta=2\pi\,,
\end{align}
where 
\begin{align}
    F_c(\omega)=\sqrt{\frac{2}{\pi}} \int\limits_0^\infty f(t) \cos(\omega t) dt
\end{align}
is the cosine Fourier transform. The sum over $m$ containing the Bessel function $K_1(\dots)$ can thus be rewritten as 
\begin{eqn}
\sum \limits_{m=1}^\infty 2 \pi m |k_\Lambda z^\Lambda|\tau_2&\cos(2\pi m k_\Lambda \zeta^\Lambda) K_1\left(2\pi m |k_\Lambda z^\Lambda| \tau_2\right)\\
&=\frac{\left(|k_\Lambda z^\Lambda| \tau_2\right)^2}{4}\sum\limits_{m=-\infty}^{\infty}\frac{1}{\left((m-k_\Lambda\zeta^\Lambda)^2+\left(|k_\Lambda z^\Lambda| \tau_2\right)^2\right)^{3/2}}-\frac{1}{2}\label{K1Poisson}
\end{eqn}
For the sum containing the Bessel function $K_0(\dots)$ we can use the Ooguri-Vafa metric close to the conifold (\ref{OoguriVafametric}) and (\ref{Conifoldmetric}) to find
\begin{align}
          \sum \limits_{m=1}^\infty\cos(2\pi m k_\Lambda \zeta^\Lambda)& K_0\left(2\pi m |k_\Lambda z^\Lambda| \tau_2\right)\label{K0Poisson}\\\nonumber &=\frac{1}{2}\sum\limits_{n=-\infty}^\infty\left[\frac{1}{\left[(|k_\Lambda z^\Lambda| \tau_2)^2+(k_\Lambda \xi^\Lambda+n)^2\right]^{1/2}}-\frac{1}{|n|}\right]+\log \left(\frac{|k_\Lambda z^\Lambda|^2\tau_2^2}{2\mu^2}\right)\,.
\end{align}
Using these expressions, we want to analyse what happens in the large volume, strong coupling limit (\ref{limit}) for which a tower of instantons has asymptotically vanishing action $\text{Re}\left(S_{D1/D(-1)}\right)=2\pi |k_\Lambda z^\Lambda| \tau_2 \lesssim e^{-\epsilon\sigma/2} \rightarrow 0$. We first focus on the $g_{z^a \bar z^b}$ component of the metric, and in particular on the term proportional to $1/\chi$ in (\ref{gzzexact}). The first term within the brackets, which originates from the classical contribution to $\chi$, scales in this limit as
\begin{align}
    \frac{\tau_2^2}{8}\mathcal{K}_{abc}t^c \sim e^{-2\sigma} \cdot \text{const.}
\end{align}
For analysing the second and third terms we will restrict to the regions of moduli space where $k_\Lambda \zeta^\Lambda \in \mathbb{Z}$, which is also the case analysed in the main text and Appedix \ref{ap:typeIIA}. In this case, the constant or logarithmic part are not the leading contribution to the  RHS of (\ref{K1Poisson}) and (\ref{K0Poisson}) anymore since now the $m=k_\Lambda \zeta^\Lambda \in \mathbb{Z}$ term in the sum dominates, which in the limit \eqref{limit} scales like 
\begin{align}
    \frac{1}{|k_\Lambda z^\Lambda|\tau_2}\sim \frac{1}{|\sum\limits_a k_a + k_0e^{-\sigma}|}e^{\sigma/2}\,. \label{scalinginZ}
\end{align}
Using the scaling behaviour \eqref{scalinginZ}, we find 
\begin{align*}
    \frac{1}{\chi}\frac{\tau_2^2}{8\pi} \sum\limits_{\bf{k}_\Lambda} n_{\bf{k}}^{(0)}\sum\limits_{m} k_a k_b K_0\left(2\pi m |k_\Lambda z^\Lambda| \tau_2\right)&\sim \frac{1}{\chi} \sum\limits_{\bf{k}_\Lambda}\frac{ n_{\bf{k}}^{(0)} k_a k_b}{|\sum\limits_a k_a + k_0e^{-\sigma}|}e^{-5/2 \sigma}\, ,\\
    \frac{1}{\chi}\frac{\tau_2^2}{16 \pi} \sum\limits_{\bf{k}_\Lambda} n_{\bf{k}}^{(0)} \sum\limits_{m}  k_a k_b  2\pi m |k_\Lambda z^\Lambda|\tau_2 K_1\left(2\pi m |k_\Lambda z^\Lambda| \tau_2\right) &\sim \frac{1}{\chi} \sum\limits_{\bf{k}_\Lambda}\frac{ n_{\bf{k}}^{(0)} k_a k_b}{|\sum\limits_a k_a + k_0e^{-\sigma}|}e^{-5/2 \sigma} \,.
\end{align*}
\noindent We are thus left with evaluating the sum over the $\bf{k}_\Lambda$. Recall from section \ref{ss:4dplanck} that the maximal allowed values for the D1/D(-1) charges of the instantons that contribute to the prepotential are given by 
\begin{align}
    k_a^\text{max} = e^{\left(\frac{1}{2}-\epsilon\right) \sigma }\,,\;\;\;\; k_0^\text{max}=e^{\left(\frac{3}{2}-\epsilon\right) \sigma }\,.
    \label{kmax}
\end{align}
We can now evaluate the sum above, which can be divided into two sums depending on whether $k_0e^{-\sigma}<\sum\limits_c k_c$ or not. If it is the case we can write  
\begin{align}
    \sum\limits_{k_c=1}^{k_c^\text{max}}\sum\limits_{k_0=1}^{\sum k_c e^{\sigma}} \frac{ n_{\bf{k}}^{(0)} k_a k_b}{|\sum\limits_c k_c + k_0 e^{-\sigma}|}\sim \exp{\left\{\left[\frac{h^{1,1}(X)+4}{2}-\left(2+h^{1,1}(X)\right)\epsilon\right]\sigma\right\}}+\dots\,,
\end{align}
where the $\dots$ stand for subleading terms in the limit $\sigma \rightarrow \infty$, and we have approximated the GV invariants to be all of the same order.\footnote{This is a very rough estimate and e.g. in the case of one-parameter CYs not the case \cite{Huang:2006hq}. However, we are interested in the contribution to the metric due to the fact that towers of instantons with different charges acquire a very small action in our limit, rather than the correction to the metric due to BPS degeneracies. As it turns out, the latter cancel out in the final expression for the metric.} The contribution to the sum coming from the sector $k_0e^{-\sigma}>\sum\limits_c k_c$ can be shown to have the same scaling behaviour. As a result, the $1/\chi$ term in \eqref{gzzexact} scales like 
\begin{align}
    \left.g_{z \bar z}\right|_{1/\chi}\sim \frac{1}{\chi}\left(-\underbrace{\exp\left(-2\sigma \right)}_{\text{class.}}+\underbrace{ \exp{\left\{\left[\frac{h^{1,1}(X)-1}{2}-\left(2+h^{1,1}(X)\right)\epsilon\right]\sigma\right\}}}_\text{inst.}\right)\,. \label{scalingchi}
\end{align}
For the $1/\chi^2$ term a similar analysis is possible, yielding
\begin{align}
    \left.g_{z \bar z}\right|_{1/\chi^2}\sim \frac{1}{\chi^2}\left(\underbrace{\exp\left(-\sigma\right)}_{\text{class.}}-\underbrace{\exp{\left\{\left[\frac{h^{1,1}(X)+1}{2}-\left(2+h^{1,1}(X)\right)\epsilon\right]\sigma\right\}}}_\text{inst.}\right)^2\,.\label{scalingchi2}
\end{align}
We thus see that in both terms in (\ref{scalingchi}) and (\ref{scalingchi2}), the instanton contribution dominates over the classical contribution if $\epsilon \le 1/2$. Recall that, given the definition \eqref{kmax}, $1/2$ is the maximum consistent value for the parameter $\epsilon$. 

To get the full scaling behaviour of the metric, we still have to include the scaling of $\chi$ as discussed in section  \ref{ss:4dplanck}. Imposing the same assumptions on the maximal D1/D(-1) charges as above, we find the scaling 
\begin{align}
    \chi \sim \exp\left\{\left(\frac{h^{1,1}(X)+3}{2}-\left(h^{1,1}(X)+1\right)\epsilon\right)\sigma \right\}\,,
    \label{chiscaling}
\end{align}
which grows as we take the limit $\sigma \rightarrow \infty$ if $\epsilon \le 1/2$. Thus the metric of the complexified K\"ahler moduli in total scales to leading order like 
\begin{align}
    g_{z \bar z}\sim \exp\left[-\left(2+\epsilon\right)\sigma\right]\,.
\end{align}

The scaling for the metric component \eqref{dchidsigma} can be deduced along the same lines. In particular one can see that in the limit \eqref{limit} this term scales as
\begin{align}
   \exp\left[-2\epsilon\sigma\right] (d\sig)^2\,.
     \label{scalingchichi}
\end{align}

With the help of these scalings of the metric components associated to the saxionic coordinates of the hypermultiplet moduli space, we can also infer the decay constants for the periodic fields.
We again use \eqref{FSmetric} to compute the the matrix of axion decay constants for the axions $c^a$ and $d_a$. The diagonal elements scale like
\begin{align}
    F_{aa}^2&\equiv \tau_2^{-2} g_{a\bar a} \sim \exp\left[\left(1-\epsilon\right)\sigma\right]\,,\\
    \tilde F_{aa}^2 &\equiv \frac{\tau_2^2}{\chi^2}g^{a\bar{a}}\sim \exp\left[-\sigma \left(h^{1,1}(X)+4-2\left(h^{1,1}(X)+\frac{3}{2}\right)\epsilon\right)\right]\,,
\end{align}
in units of the corrected Planck mass. As before, one can derive these results from a direct computation of the metric, as we discuss in the Appendix \ref{ap:typeIIA}. The same analysis can be used to compute the decay constants for the remaining axions which, unlike the others, do not fit into the metric Ansatz \eqref{FSmetric}. One instead finds that
\begin{align}
   F_{00}^2&\sim \exp\left[\left(3-\epsilon\right)\sigma\right]\,, \\
    \tilde F_{00}^2&\sim \exp\left[-\sigma \left(h^{1,1}(X)+6-2\left(h^{1,1}(X)+\frac{3}{2}\right)\epsilon\right)\right]\,,
\end{align}
with the first and second lines being the decay constants for $\tau_1$ and $c^0$, respectively.


\section{Instantons in type IIA CY compactifications}
\label{ap:typeIIA}
In this work we have mostly focused on the type IIB hypermultiplet moduli space of a CY $X$. However, a similar reasoning can be applied to the hypermultiplet moduli space of type IIA compactified on the CY $Y$, mirror dual to $X$. The hypermultiplet moduli space of type IIA on $Y$ contains, besides the complex structure moduli of the $Y$, the RR 3-form axions obtained by reducing the IIA 3-form $C_3$ along 3-cycles. To describe the  complex structure moduli space, define an integral symplectic basis of 3-cycles $(A_\Lambda, B^\Lambda)$, $\Lambda=0,\dots h^{2,1}(Y)$, satisfying 
\begin{align}
    A^I\cdot B_J=-B_J\cdot A^I =\delta^I_J\,.
\end{align}
The coordinates of the complex structure moduli space are now defined via the periods of the holomorphic 3-form $\Omega$ of $Y$ over these 3-cycles
\begin{align}
    F_\Lambda = \int\limits_{B^\Lambda} \Omega\,,\;\;\; X^\Lambda = \int\limits_{A_\Lambda} \Omega\,,\;\;\; z^\Lambda=\frac{X^\Lambda}{X^0}\,.
\end{align}
In the same basis of 3-cycles, the 3-form axions are given by
\begin{align}
    \zeta^\Lambda = \int\limits_{A_\Lambda} C_3 \,,\;\;\;\;\; \tilde\zeta_\Lambda = \int\limits_{B^\Lambda} C_3\,.
\end{align}
Applying the mirror map to the $SL(2,\mathbb{Z})$-completed expression for $\chi$ in the IIB hypermultiplet moduli space, one obtains the corresponding expression for the IIA hypermultiplet moduli space \cite{Alexandrov:2008gh}
\begin{align}
    \chi^\text{IIA}=\frac{\mathcal{R}^2}{4}K(z,\bar z) +\frac{\chi_Y}{192\pi} +\frac{\mathcal{R}}{8\pi^2}\sum\limits_\gamma \Omega_\gamma \sum\limits_{m>0} \frac{|Z_\gamma|}{m} \cos\left(2\pi m \Theta_\gamma\right)K_1\left(4\pi m \mathcal{R}|Z_\gamma|\right)\,, 
    \label{chiIIA}
\end{align}
where the coordinate $\mathcal{R}$ is the mirror dual of the IIB dilaton $\tau_2$ and $K(z,\bar z)=2\text{Im}\left(z^\Lambda \bar F_\Lambda\right)$ is the classical K\"ahler potential in complex structure moduli space. Moreover, $\chi_Y=-\chi_X$ is the Euler characteristic of the mirror $Y$ of $X$ and we have defined 
\begin{align}
    Z_\gamma\equiv & \left(k_\Lambda z^\Lambda -l^\Lambda F_\Lambda\right)\,,\\
    \Theta_\gamma\equiv& k_\Lambda \zeta^\Lambda - l^\Lambda \tilde \zeta_\Lambda\,.
\end{align}
Finally, the instanton measure $\Omega_\gamma$ counts the number of special Lagrangian 3-cycles homologous to $k_\Lambda A^\Lambda$ in $H_3(Y,\mathbb{Z})$. 

The large volume/strong coupling limit analysed in the type IIB setup now translates into the large complex structure limit, as can easily be seen by looking at $t^a\equiv \text{Im}z^a$ in the limit 
\begin{align}
    t^a \rightarrow e^\sigma t_0^a\,,\;\;\;\;\mathcal{R}\sim e^{-3\sigma/2} \mathcal{R}(t_0)\,,\;\;\; \sigma\rightarrow +\infty \,. \label{limitIIA}
\end{align}
For $l^\Lambda=0$, i.e. the case where the D2-brane instantons only wrap A-cycles, the instanton action thus scales like 
\begin{align}
    \mathcal{R}|Z_\gamma|\sim e^{-\frac{\sigma}{2}}\left(\frac{\sum k_a}{t_0^{1/2}}\right) +e^{\frac{-3\sigma}{2}}\left(\frac{k_0}{t_0^{3/2}}\right)\,.
\end{align}
Hence, we identify our case to be correspond to $d=3$ in the language of \cite{Grimm:2018ohb} and we see that, as in the IIB case, there are towers of instantons which have an asymptotically vanishing action.

The metric for the type IIA hypermultiplet moduli space for the case of purely electrically charged D2-brane instantons, i.e. $l^\Lambda=0$, has already been calculated in \cite{Alexandrov:2014sya}. In the following we review their result and analyse the behaviour of the metric in the particular limit \eqref{limitIIA}. As in the IIB case in the main text, we consider a path on the hypermultiplet moduli space along which the vevs for the periodic coordinates vanish, which simplifies the analysis of the metric significantly.

In general, the function $\chi^\text{IIA}$ can be used to calculate the metric on the so-called twistor space $\mathcal{Z}$ associated to the hypermultiplet moduli space $\mathcal{M}_{\rm HM}$. $\mathcal{Z}$ is a $\mathbb{C}P^1$ bundle over $\mathcal{M}_{\rm HM}$ with connection given by the SU(2) part of the the Levi-Civita connection on $\mathcal{M}_{\rm HM}$. The K\"ahler-Einstein metric for the twistor space can be written as
\begin{align}
    ds_\mathcal{Z}^2=\frac{|D\mathbf{t}|^2}{\left(1+\mathbf{t} \mathbf{\bar t}\right)^2}+\frac{\nu}{4}ds_{\mathcal{M}_{\rm HM}}^2 \label{eq::twistormetric}
\end{align}
where $\mathbf{t}$ is a complex coordinate on $\mathbb{C}P^1$ and $\nu$ sets the curvature of $\mathcal{M}_{\rm HM}$. The 1-form $D\mathbf{t}$ is given by
\begin{align}
    D\mathbf{t}=d\mathbf{t}+p_+-ip_3\mathbf{t}+p_-\mathbf{t}^2\,,
\end{align}
where the $p_*$ are the components of the SU(2) connection on $\mathcal{M}$. The K\"ahler potential for the twistor space reads 
\begin{align}
    K_\mathcal{Z}=\log \frac{1+\mathbf{t} \mathbf{\bar t}}{|\mathbf{t}|}+ \text{Re}\, \log\chi^{IIA}(x^\mu, \mathbf{t})\,.
\end{align}
In \cite{Alexandrov:2014sya} the metric is however not calculated directly from $\chi^{IIA}$, but by exploiting the SU(2) connection underlying the quaternionic K\"ahler geometry which is given by
\begin{eqn}
    p_+&=-\frac{i}{4r} \frac{\mathcal{R}}{8\pi^2} \sum\limits_\gamma n_\gamma Z_\gamma d J_\gamma^{(1)}\,,\\
    p_3&=\frac{1}{8r}\left[2\mathcal{R}^2K\left(N_a dz^a -N_{\bar a}d\bar z^a\right)-\frac{\mathcal{R}}{4\pi^2}\sum \limits_\gamma \Omega_\gamma \left(\mathcal{J}_\gamma^{(1,+)}d Z_\gamma -\mathcal{J}_\gamma^{(1,-)}d\bar Z_\gamma\right)\right]\,,\\
    p_-&=\left(p_+\right)^*\,. \label{eq::ps}
\end{eqn}
Here, the functions $\mathcal{J}_\gamma^{(\star, \star)}$ are twistorial integrals given by 
\begin{eqn}
    \mathcal{J}_\gamma^{(1)}&=\int_{l_\gamma}\frac{dt}{t}\log\left(1-\sigma_\gamma e^{-2\pi i \Xi_\gamma(t)}\right)\,,\;\;\;\;\;\;\;\;\;\;\;\; \mathcal{J}_\gamma^{(2)}=\int_{l_\gamma}\frac{dt}{t}\frac{1}{\sigma_\gamma e^{-2\pi i \Xi_\gamma(t)} -1}\\
    \mathcal{J}_\gamma^{(1,\pm)}&=\pm\int_{l_\gamma}\frac{dt}{t^{1\pm 1}}\log\left(1-\sigma_\gamma e^{-2\pi i \Xi_\gamma(t)}\right)\,,\;\;\; \mathcal{J}_\gamma^{(2,\pm)}=\pm\int_{l_\gamma}\frac{dt}{t^{1\pm 1}}\frac{1}{\sigma_\gamma e^{-2\pi i \Xi_\gamma(t)} -1} \,,\label{eq::js}
\end{eqn}
where $l_\gamma$ is a ``BPS ray" on $\mathbb{C}P^1$ and $\sigma_\gamma=\pm1$ which, for instantons with only electric charges, can be chosen to be +1. Note that the functions on the right are the derivatives of the funcitons on the left. Furthermore, we introduced $\Xi_\gamma$ which in our case (vanishing vevs for the axion fields) is given by 
\begin{align}
    \Xi_\gamma =\mathcal{R}\left(t^{-1}Z_\gamma - t\bar Z_\gamma \right)\,. 
\end{align}
Next, we note that our expression for $\chi^\text{IIA}$ actually corresponds to the type IIA 4d dilaton $r$ which in terms of the $\mathcal{J}_\gamma$ functions can be written as
\begin{align}
    r=\frac{\mathcal{R}^2}{4}K+\frac{\chi_Y}{192\pi}-\frac{i\mathcal{R}}{32 \pi^2}\sum\limits_\gamma \Omega_\gamma\left(Z_\gamma \mathcal{J}_\gamma^{(1,+)}+\bar Z_\gamma \mathcal{J}_{\gamma}^{(1,-)}\right)\,. \label{eq::rr}
\end{align}
Comparing this expression with \eqref{chiIIA}, we can express the functions $\mathcal{J}_\gamma$ in terms of sums over Bessel functions:
\begin{eqn}
    \mathcal{J}_\gamma^{(1,+)}=4i\sqrt{\frac{\bar Z_\gamma }{Z_\gamma}}\sum\limits_{m>0} \frac{1}{m} &\cos\left(2\pi m \Theta_\gamma\right)K_1\left(4\pi m \mathcal{R}|Z_\gamma|\right)\,,\;\;\; \mathcal{J}_\gamma^{(1,-)}=\frac{Z_\gamma}{\bar Z_\gamma}\mathcal{J}_\gamma^{(1,+)}\,. \label{eq:identification}
\end{eqn}
This identification can also be seen by using the representation of the Bessel function
\begin{align}
    \int\limits_0^\infty\frac{dt}{t}\left(\alpha t+\frac{\beta}{t}\right)e^{-\frac{1}{2}\left(\alpha t+\frac{\beta}{t}\right)}=4\sqrt{\alpha \beta}K_1\left(\sqrt{\alpha \beta}\right)\,,\label{eq::Besselidentity}
\end{align}
and using that $\log\left(1-x\right)=-\sum\limits_{m=1}^\infty \frac{x^m}{m}$. In particular we can rewrite the instanton part of \eqref{eq::rr} as
\begin{eqn*}
    -\frac{i\mathcal{R}}{32 \pi^2}\sum\limits_\gamma \Omega_\gamma\left(Z_\gamma \mathcal{J}_\gamma^{(1,+)}+\bar Z_\gamma \mathcal{J}_{\gamma}^{(1,-)}\right)=\frac{1}{32\pi^3}\int\limits_0^\infty\frac{dt}{t}\sum\limits_{m=1}^\infty\frac{1}{4m^2}\left(\alpha_mt+\frac{\beta_m}{t}\right)e^{-\frac{1}{2}\left(\alpha_mt+\frac{\beta_m}{t}\right)}\,,
\end{eqn*} 
where for simplicity we have set $\Theta_\gamma=0$ -- which is also the case considered below -- and defined
\begin{align*}
    \alpha_m=-4\pi\mathcal{R}m \bar Z_\gamma\,,\;\;\;\; \beta_m=4\pi i \mathcal{R} m Z_\gamma \,. 
\end{align*}
Using \eqref{eq::Besselidentity}, we then see that
\begin{align}
     -\frac{i\mathcal{R}}{32 \pi^2}\sum\limits_\gamma \Omega_\gamma\left(Z_\gamma \mathcal{J}_\gamma^{(1,+)}+\bar Z_\gamma \mathcal{J}_{\gamma}^{(1,-)}\right)=\frac{1}{32\pi^3}\sum\limits_{m=1}^\infty\sqrt{\alpha_m\beta_m}K_1\left(\sqrt{\alpha_m\beta_m}\right)\,,
\end{align}
which reduces to the expression in \eqref{chiIIA} upon inserting $\alpha_m$ and $\beta_m$. \\
To find the metric on the hypermultiplet moduli space it suffices to know the triplet of quaternionic 2-forms on the quaternionic manifold. This can be expressed through the SU(2) connection as
\begin{align}
    \vec \omega = - 2\left(d\vec p +\frac{1}{2}\vec p\times \vec p\right)\,, 
\end{align}
and so $\omega^3$ is for instance given by
\begin{align}
    \omega^3=-2 dp_3 + 4i p_+\wedge p_-\,. \label{eq::omega3general}
\end{align}
For a given compatible almost complex structure $J^3$, the metric can be calculated from
\begin{align}
    g(X,Y)=\omega^3(X, J^3 Y)\,. \label{eq::metrickahlerform}
\end{align}
This almost complex structure can be specified by choosing a basis of (1,0)-forms. A suitable set in this case is given by \cite{Alexandrov:2014sya}
\begin{eqn}
    dz^a&\,,\\
    \mathcal{Y}_\Lambda=&\, d \tilde \zeta_\Lambda - F_{\Lambda \Sigma} d \zeta^\Sigma -\frac{1}{8\pi^2} \sum\limits_\gamma \Omega_\gamma q_\Lambda d \mathcal{J}_\gamma^{(1)}\,,\\
    \Sigma =& \, dr +2 c\, d \log \mathcal{R } - \frac{i}{16\pi^2} \sum\limits_\gamma \Omega_\gamma \left(\mathcal{R}Z_\gamma d \mathcal{J}_\gamma ^{(1,+)}-\mathcal{J}_\gamma ^{(1,-)}d\left(\mathcal{R}\bar Z_\gamma\right)\right)\\
    &+\frac{i}{4}\left(d\sigma -\zeta^\Lambda d \tilde \zeta_\Lambda +\tilde \zeta_\Lambda d \zeta^\Lambda\right)\,.\label{eq::complexstructure}
\end{eqn}
 Using \eqref{eq::omega3general} and \eqref{eq::ps} as well as the basis of (1,0)-forms, one can show that $\omega^3$ can be written as (cf. eq. (B.37) of \cite{Alexandrov:2014sya})
\begin{eqn}
    \omega^3=&\; i\frac{\hat\Sigma \wedge \hat{\bar \Sigma} }{4r^2\left(1-\frac{2r}{R^2U}\right)}+\frac{i\mathcal{R}^2}{4r^2}z^\Lambda \bar z^\Sigma \mathcal{Y}_\Lambda \wedge \overline{\mathcal{Y}}_\Sigma \\
    &-\frac{i\mathcal{R}^2}{2r}N_{\Lambda \Sigma}dz^\Lambda \wedge d\bar z ^\Sigma -\frac{i\mathcal{R}^2}{8\pi r}\sum\limits_\gamma \Omega_\gamma \mathcal{J}_\gamma^{(2)}q_\Lambda q_\Sigma d\bar z^\Lambda \wedge dz^\Sigma \\
    &+\frac{iK}{rU} \left[\frac{\mathcal{R}^2K}{4}\left(N_a dz^a+N_{\bar b}d\bar z^b\right)+\frac{\mathcal{R}^2}{8\pi}\sum\limits_\gamma\Omega_\gamma \mathcal{J}_\gamma^{(2)}d|Z_\gamma|^2)\right]\\
    &\;\;\;\;\;\;\;\;\;\;\;\;\wedge\left(N_a dz^a-N_{\bar b}d\bar z^b\right)\, + \dots \label{eq::omega3}
\end{eqn}
where a new (1,0)-form $\hat \Sigma$ has been introduced, which in terms of the above basis  reads
\begin{align}
   \hat \Sigma =\Sigma +f_\Lambda dz^\Lambda +g^\Lambda \mathcal{Y}_\Lambda\,,
\end{align}
with the functions $f_\Lambda$ and $g_\Lambda$ such that \eqref{eq::omega3} reproduces \eqref{eq::omega3general}. Finally, $N_a=\partial_{z^a} \log K$ and the dots stand for terms that are quadratic in the periodic directions, and which do not play any role in computing the line element along the trajectory \eqref{limitIIA}. 

In fact, since we are just interested in the directions in which the RR-fields do not change, we may project the above (1,0)-forms onto their components corresponding to complex structure and dilaton directions.
The corresponding components of $\hat \Sigma$ are
\begin{eqn}
    \hat \Sigma \supset\left(2-\frac{2r}{\mathcal{R}^2U}\right)dr +&\frac{i}{4}\left[2\mathcal{R}^2K \left(1-\frac{4r}{\mathcal{R}^2U}\right)\left(N_a dz^a-N_{\bar b}d\bar z^b\right)\right.\\
    &\left.-\frac{\mathcal{R}}{4\pi^2}\sum\limits_\gamma \Omega_\gamma\left(\mathcal{J}_\gamma^{(1,+)}dZ_\gamma - \mathcal{J}_\gamma^{(1,-)}d\bar Z_\gamma \right)\right]\,. 
\end{eqn}
Here $U$ is given by \cite{Alexandrov:2014sya} 
\begin{eqn}
    U=K-\frac{1}{2\pi}\sum \limits_\gamma \Omega_\gamma |Z_\gamma|
    ^2 \mathcal{J}_\gamma^{(2)}+v_\Lambda M^{\Lambda \Sigma }\bar v_\Sigma \,,
\end{eqn}
where $M^{\Lambda \Sigma}$ is the inverse matrix of 
\begin{eqn}
    M_{\Lambda \Sigma}&=N_{\Lambda \Sigma }-\sum\limits_\gamma \Omega_\gamma \mathcal{J}_\gamma ^{(2)}q_\Lambda q_\Sigma \,, \label{eq::MLAMBDASIGMA}
\end{eqn}
and the vector $v_\Lambda$ is given by 
\begin{eqn}
    v_\Lambda&=\frac{1}{4\pi}\sum \limits_\gamma \Omega _\gamma q_\Lambda\left(Z_\gamma \mathcal{J}_\gamma^{(2,+)}+\bar Z_\gamma \mathcal{J}_\gamma ^{(2,-)}\right) \,. 
\end{eqn}
Reading off the metric for the non-periodic directions from \eqref{eq::omega3}, one finds that it can be split into four parts
\begin{align}
    ds^2=ds^2_1+ds^2_2 +ds^2_3+ds^2_4\,.
    \label{fourds}
\end{align}
The first one is given by
\begin{eqn}
    ds^2_1=&\frac{\left(1-\frac{2r}{\mathcal{R}^2U}\right)}{2r^2}dr^2+\frac{\mathcal{R}^4K^2 \left(1-\frac{4r}{\mathcal{R}^2U}\right)^2}{4r^2\left(1-\frac{2r}{\mathcal{R}^2U}\right)}N_aN_{\bar b}dz^a d\bar z^b \\
    +&\left\{\frac{\mathcal{R}^3 K\left(1-\frac{4r}{\mathcal{R}^2U}\right)}{32\pi^2r^2\left(1-\frac{2r}{\mathcal{R}^2U}\right)}\sum \limits_\gamma \Omega _\gamma \left(q_{ b}N_a  \mathcal{J}_\gamma^{(1,-)}+ q_aN_{\bar b} \mathcal{J}_\gamma^{(1,+)}\right)\right. \\ 
    &+\frac{1}{32r^2\left(1-\frac{2r}{\mathcal{R}^2U}\right)}\frac{\mathcal{R}^2}{16\pi^4}\left(\sum \limits_\gamma \Omega _\gamma q_a\mathcal{J}_\gamma^{(1,+)}\right)\left(\sum \limits_\gamma \Omega _\gamma q_{ b}\mathcal{J}_\gamma^{(1,-)}\right)\\
    &\left.+\frac{\mathcal{R}^4}{4r^2}\left|\frac{1}{8\pi^2}\left(\sum \limits_\gamma \Omega _\gamma q_a Z_\gamma\left(\mathcal{J}_\gamma^{(2,-)}+\mathcal{J}_\gamma^{(2,+)}\right)\right)\right|^2\right\}dz^ad\bar z^b\,,\label{eq::ds1}
\end{eqn}
which contains the metric along the $r$ direction and the $dz^a d\bar z ^b $ components that go with $1/r^2$. The next part of the metric contains all the terms that go like $1/r$: 
\begin{align}
    ds^2_2=-\frac{\mathcal{R}^2}{r}\left(N_{\Lambda \Sigma} - \frac{1}{4\pi} \sum\limits_\gamma \Omega_\gamma \mathcal{J}_\gamma^{(2)} q_\Lambda q_\Sigma \right) dz^\Lambda d\bar z^\Sigma \,. 
\end{align}
The third part of the metric comes from the last term in \eqref{eq::omega3} and includes terms that go like $1/(rU)$: 
\begin{eqn}
    ds^2_3=\frac{K\mathcal{R}^2}{rU}\left(KN_a N_{\bar b}+\frac{1}{4\pi}\sum\limits_\gamma  \Omega_\gamma \mathcal{J}_\gamma^{(2)} \left[Z_\gamma q_b N_a+\bar Z_\gamma q_aN_{\bar b}\right] \right)dz^a d\bar z^b \label{eq::ds3}
\end{eqn}
The last contribution to \eqref{fourds} originates from the term in \eqref{eq::omega3} containing $\mathcal{Y}_\Lambda$ and the fact that the differential $d\mathcal{J}_\gamma^{1}$ has a $d\mathcal{R}$ component. Thus, this part of the metric involves terms like $d\mathcal{R}^2$ and $d\mathcal{R}dz$ and it is given by
\begin{eqn}
    ds^2_4=&\frac{\mathcal{R}^3d\mathcal{R}}{32\pi r^2}\left(\sum\limits_\gamma \Omega_\gamma Z_\gamma q_a\left[\mathcal{J}_\gamma^{(2,+)}dz^a +\mathcal{J}_\gamma^{(2,-)}d\bar z^a\right]\right)
     \left(\sum\limits_\gamma \Omega_\gamma \bar Z_\gamma\left[ \mathcal{\bar J}_\gamma^{(2,+)}\bar Z_\gamma + \mathcal{\bar J}_\gamma^{(2,-)}Z_\gamma\right]\right)\\
    &+\frac{\mathcal{R}^2}{32 r^2}\left|\sum\limits_\gamma \Omega_\gamma Z_\gamma \left(Z_\gamma \mathcal{J}_\gamma^{(2,+)}+\bar Z_\gamma \mathcal{J}_\gamma^{(2,-)}\right)\right|^2 d\mathcal{R}^2+ c.c.\,. \label{eq::ds4}
\end{eqn}
We thus see that all terms of the metric including $d\mathcal{R}$ go like $1/r^2$ unlike the terms in $ds_2^2$ that go like $1/r$. \\

Let us now compare the metric obtained above with the expressions that we get in appendix \ref{ap:metric}, where we adapted the FS metric Ansatz \eqref{FSmetric} to the quantum corrected,  SL(2,$\mathbb{Z}$)-completed K\"ahler potential $\chi$ in the mirror dual IIB hypermultiplet moduli space. For this we will use that in the presence of D(-1)/D1-brane instantons, the quantum corrected mirror map coincides with the classical one except for  the axions $\tilde{\zeta}_\Lam$ \cite{Alexandrov:2009qq}, whose vev is however set to zero in the following. 

Let us first focus on the metric components along $z^a$, which correspond to \eqref{gzzexact}. We start with the part of the metric $ds^2_2$, which goes like $1/r$ and consists of a classical and a quantum part given by the first and second term respectively. The analog of the first term in the expression \eqref{gzzexact} is just the classical part of the $1/\chi$ contribution and we identify 

\begin{center}
\begin{tabular}{c|c}
    IIA exact result & IIB K\"ahler potential approximation \\ \hline 
    $\frac{\mathcal{R}^2}{r} N_{\Lambda \Sigma}$ & $\frac{\tau_2^2}{8\chi}{\cal K}_{ab}$
\end{tabular}
\end{center}

\noindent The second part of $ds^2_2$ appears when taking the second derivatives of $r$ w.r.t. $z^a$ and $\bar z^b$ and we thus can identify 

\begin{center}
\begin{tabular}{cc|cc}
    IIA exact result &&& IIB K\"ahler potential approximation \\ \hline 
   $\frac{\mathcal{R}^2}{4\pi r} \sum\limits_\gamma \Omega_\gamma \mathcal{J}_\gamma^{(2)} q_a q_b $&& &$\frac{\tau_2^2}{8\pi \chi} \sum_{{\bf k_\Lambda}} n_{\bf{k}}^{(0)}\sum_{m} k_a k_b K_0\left(2\pi m |k_\Lambda z^\Lambda| \tau_2\right)$
\end{tabular}
\end{center}

\noindent We saw in appendix \ref{ap:metric} that the IIB expression above is the leading contribution to the metric in our limit. Thus, to leading order, the approximation for the metric in the IIB case in appendix \ref{ap:metric} has the same behaviour as the exact result in the IIA case, as long as the terms in $ds^2_1$ and $ds^2_3$ of the exact metric are not dominant over the contributions in $ds_2^2$. To see this, we find the analogues of these terms in \eqref{gzzexact}. 

Let us start with noting that in our limit to leading order we have the scaling $U\sim r/\mathcal{R}^2$. Looking at \eqref{eq::ds1} we  see that
\begin{align}
    \frac{r}{\mathcal{R}^2U}\sim \mathcal{O}(1)\,. 
\end{align}
Thus indeed all the terms in \eqref{eq::ds1} scale like $1/r^2$ up to corrections that become negligible in our limit, and in particular also the terms in \eqref{eq::ds3} scale like $1/r^2$. Up to the corrections induced by $U$, the second term in \eqref{eq::ds1} and the first term in \eqref{eq::ds3} can be identified as the classical contribution to the metric. We can furthermore identify them with the classical contribution to the $1/\chi^2$ term in \eqref{gzzexact}: 

\begin{center}
\begin{tabular}{cc|cc}
    IIA exact result &&& IIB K\"ahler potential approximation \\ \hline 
   $\frac{K^2 \mathcal{R}^4}{r^2} N_aN_{\bar b} $&& &$\frac{\tau_2^4 K^2}{64\chi^2}\, \mathcal{K}_{a}\mathcal{K}_{b}$
\end{tabular}
\end{center}

\noindent  The terms on the left side can be obtained from the product of the first-derivatives of the classical expression for $\log r$ with respect to $z^a$ and $\bar z^b$, respectively. Taking the quantum-corrected expression for $r$ in \eqref{eq::rr}, we see that we get the term in the brackets of \eqref{eq::ds1} and the second term in \eqref{eq::ds3} which we can thus identify with the corresponding quantum corrections of \eqref{gzzexact} to the $1/\chi^2$ term, e.g.

\begin{center}
\begin{tabular}{c|c}
    IIA exact result & IIB K\"ahler potential approximation \\ \hline &\\
   $\frac{K\mathcal{R}^4}{r^2}\frac{1}{4\pi}\sum\limits_\gamma  \Omega_\gamma \mathcal{J}_\gamma^{(2)} Z_\gamma q_b N_a$ &$\frac{\tau_2^4}{16 \pi \chi^2}\mathcal{K}_{a} \sum\limits_{{\bf k_\Lambda}} n_{\bf{k}}^{(0)}  \sum_{m} k_\Lambda z^\Lambda k_b K_0\left(2\pi m |k_\Lambda z^\Lambda| \tau_2\right) $\\ &\\\hline & \\
   $\frac{\mathcal{R}^4}{4r^2}\left|\frac{1}{8\pi^2}\left(\sum \limits_\gamma \Omega _\gamma q_a Z_\gamma \left(\mathcal{J}_\gamma^{(2,-)}+\mathcal{J}_\gamma^{(2,+)}\right)\right)\right|^2$ & $\frac{\tau_2^4}{64\pi^2\chi^2} \left|\sum\limits_{{\bf k_\Lambda}} n_{\bf{k}}^{(0)}  \sum_{m} k_\Lambda \bar z^\Lambda k_a K_0\left(2\pi m |k_\Lambda z^\Lambda| \tau_2\right)\right|^2$ \\&\\ 
  
\end{tabular}
\end{center}

\noindent The remaining terms can be identified accordingly. Note that in the IIB K\"ahler approximation we set the periodic coordinates to zero in comparison to \eqref{gzzexact}. 

We thus see that, in this limit, the contribution from $ds_2^2$ to the metric is dominant over the others, which correspond to sub-leading terms in appendix \ref{ap:metric}. Notice that, as far as the $z$ components of the metric are considered, the results of appendix \ref{ap:metric} carry over to the leading order of the exact metric in type IIA.

Having analysed the metric along the coordinates $z^a$, we can now turn to the 4d dilaton. Recall that in Appendix \ref{ap:metric} we promoted the Ansatz \eqref{FSmetric}, for which the 4d dilaton and the other moduli decouple, to the quantum corrected regime. There, an important contribution to the line element along the trajectory \eqref{limit} arises from \eqref{dchidsigma}. In the exact type IIA metric, the analogous term is given by the first term in \eqref{eq::ds1} which in the IIA limit \eqref{limitIIA} reduces to $dr^2/r^2$. Evaluating this term in our limit thus gives rise to the same behaviour as displayed in \eqref{scalingchichi}.

However, in the IIA exact metric there are also the terms in \eqref{eq::ds4}, which give us an additional contribution for the metric components involving the coordinate $\cal{R}$ dual to the 10d type IIB dilaton. Since the limit \eqref{limitIIA} is defined via the scaling of $\mathcal{R}$, it is important to analyze the behaviour of these contributions, to make sure that they do not dominate over the part that is already present in the Ansatz \eqref{FSmetric}. The functional dependence of the terms in \eqref{eq::ds4} suggests that they have the same behaviour as $dr^2/r^2$, since they can be written as the pull-back of $\log r$:
\begin{eqn}
    &\frac{\mathcal{R}^2}{32 r^2}\left|\sum\limits_\gamma \Omega_\gamma Z_\gamma \left(Z_\gamma \mathcal{J}_\gamma^{(2,+)}+\bar Z_\gamma \mathcal{J}_\gamma^{(2,-)}\right)\right|^2d\mathcal{R}^2\sim \left(\frac{\partial \log r}{\partial \cal{R}} \right)^2 d\mathcal{R}^2 \\
    &\frac{\mathcal{R}^3d\mathcal{R}}{32\pi r^2}\left(\sum\limits_\gamma \Omega_\gamma Z_\gamma q_a \mathcal{J}_\gamma^{(2,+)}dz^a\right)
     \left(\sum\limits_\gamma \Omega_\gamma \bar Z_\gamma\left[ \mathcal{\bar J}_\gamma^{(2,+)}\bar Z_\gamma + \mathcal{\bar J}_\gamma^{(2,-)}Z_\gamma\right]\right)\\&\sim \left(\frac{\partial \log r}{\partial \mathcal{R}}\frac{\partial \log r}{\partial z^a}\right)d\mathcal{R} dz^a\,. 
\end{eqn}
Therefore in the limit \eqref{limitIIA} these terms scale similarly to  $dr^2/r^2$. We can then conclude that the Ansatz \eqref{FSmetric} also captures the leading behaviour of the exact metric in the limit \eqref{limitIIA} along the direction $\mathcal{R}$. 

So far, we have only investigated the metric components for the dilaton and the complex structure moduli $z^a$. In the main text, we are also interested in the metric components quadratic in the RR fields since, classically, these correspond to the decay constants of the corresponding axions. We first focus on the RR fields $ \zeta^\Lambda$ arising from reducing the IIA 3-form along the A-cycles. In this case, we directly use the expression for the metric found in \cite{Alexandrov:2014sya}. This metric contains to leading order two terms quadratic in $d \zeta^\Lambda$ that are induced fully by the instanton corrections. The first term arises from
\begin{align}
  \frac{\mathcal{R}^2}{2r^2}\left|z^\Lambda \mathcal{Y}_\Lambda\right|\supset \frac{\mathcal{R}^2}{32r^2}\left(\sum\limits_\gamma \Omega_\gamma Z_\gamma \mathcal{J}_\gamma^{(2)}q_\Lambda d \zeta^\Lambda \right)\left(\sum\limits_\gamma \Omega_\gamma \bar Z_\gamma  \mathcal{\bar J}_\gamma^{(2)}q_\Sigma d \zeta^\Sigma \right)  \,,
\end{align}
and the second term is given by
\begin{align}
    -\frac{1}{r}M^{\Lambda \Sigma}\mathcal{Y}_\Lambda  \mathcal{\bar Y}_\Sigma\supset -\frac{1}{r}M^{\Lambda \Sigma}\left(\sum\limits_\gamma \Omega_\gamma \mathcal{J}_\gamma^{(2)}q_\Lambda q_\Pi d \zeta^\Pi \right)\left(\sum\limits_\gamma \Omega_\gamma \mathcal{\bar J}_\gamma^{(2)}q_\Sigma q_{\Pi'}d \zeta^{\Pi'} \right) \,,  \label{eq::RRfields2}
\end{align}
where we recall that $M^{\Lambda \Sigma}$ is the inverse of \eqref{eq::MLAMBDASIGMA} which in our limit is dominated by the instanton contribution. Since we are only interested in the scaling behaviour of the metric component quadratic in the RR fields, we can approximate the above expression by
\begin{align}
   \eqref{eq::RRfields2}\sim \frac{1}{r} \sum\limits_\gamma \Omega_\gamma q_\Lambda q_{\Sigma}\mathcal{J}_\gamma^{(2)} d \zeta^{\Lambda }d \zeta^{\Sigma }\,.
\end{align}
Thus comparing with the result for the metric components quadratic in the K\"ahler moduli, we see that the metric $g_{ \zeta  \zeta}$ behaves to leading order in our limit like 
\begin{align}
    g_{ \zeta^\Lambda  \zeta^\Sigma } \sim \frac{1}{\mathcal{R}^2} g_{z^\Lambda \bar z^\Sigma}\,. \label{gzetazeta}
\end{align}
Here, $g_{z^\Lambda \bar z^\Sigma}$ reduces to the leading contribution of $g_{z^a \bar z^b}$ for $(\Lambda,\Sigma)=(a,b)$ and in analogy to the leading expression for $g_{z^a \bar z^b}$ we further defined the components $g_{z^0 \bar z^0}\equiv \frac{\mathcal{R}^2}{4\pi r} \sum\limits_\gamma \Omega_\gamma \mathcal{J}_\gamma^{(2)} q_0 q_0$ and similar for $g_{z^0 \bar z^a}$  even though they do not correspond to metric components for physical field. We note that \eqref{gzetazeta} is reminiscent of the expression for the components quadratic in the RR field of the classical IIB hypermultiplet moduli space metric given by \eqref{FSmetric}. 

From the term on the lhs of \eqref{eq::RRfields2} we can also infer the behaviour of the decay constants of the axions $\tilde \zeta_\Lambda$ obtained by reducing the IIA 3-form along the B-cycles. We obtain to leading order in our limit 
\begin{align}
    -\frac{1}{r}M^{\Lambda \Sigma}\mathcal{Y}_\Lambda  \mathcal{\bar Y}_\Sigma\supset -\frac{1}{r}M^{\Lambda \Sigma}d \tilde \zeta_\Lambda d\tilde \zeta_\Sigma \sim \left[\frac{r}{2\pi }\sum\limits_\gamma\Omega_\gamma \mathcal{J}_\gamma^{(2)}q_\Lambda q_\Sigma \right]^{-1}d \tilde \zeta_\Lambda d\tilde \zeta_\Sigma \sim \frac{\mathcal{R}^2 g^{z^\Lambda \bar z^\Sigma}}{r^2} d\tilde \zeta_\Lambda d\tilde \zeta_\Sigma\,,
\end{align}
where we used in the last step that the term in bracket resembles the leading contribution to the metric component quadratic in the complexified K\"ahler moduli. Here, $g^{z^\Lambda \bar z^\Sigma}$ is to be understood as the inverse of the $g_{z^\Lambda \bar z^\Sigma}$ defined above. Again the above result shows that the quantum corrected decay constants for the $\tilde \zeta_\Lambda$ have, to leading order, a similar dependence on $g_{z \bar z}$ as the classical decay constants in \eqref{FSmetric}.

To sum up, in the IIA case we also have towers of instantons becoming very relevant due to their small action in the large complex structure limit. By directly computing the quantum corrected metric, one finds that its leading behaviour reproduces the one obtained in Appendix \ref{ap:metric} by adapting the classical metric Ansatz \eqref{FSmetric} to the type IIB $SL(2,\mathbb{Z})$-invariant 4d dilaton. As such, the analysis of the main text applies to this case as well, and the classical infinite distance along \eqref{limitIIA} is resolved by quantum corrections.


\begin{thebibliography}{10}


\bibitem{Vafa:2005ui} 
  C.~Vafa,
  {\em ``The String landscape and the swampland,''}
  hep-th/0509212.

\bibitem{Brennan:2017rbf} 
  T.~D.~Brennan, F.~Carta and C.~Vafa,
  {\em ``The String Landscape, the Swampland, and the Missing Corner,''}
  PoS TASI {\bf 2017}, 015 (2017)
  [arXiv:1711.00864 [hep-th]].

\bibitem{Palti:2019pca}
  E.~Palti,
  {\em ``The Swampland: Introduction and Review,''}
  arXiv:1903.06239 [hep-th].

\bibitem{ArkaniHamed:2006dz} 
  N.~Arkani-Hamed, L.~Motl, A.~Nicolis and C.~Vafa,
  {\em ``The String landscape, black holes and gravity as the weakest force,''}
  JHEP {\bf 0706}, 060 (2007)
  [hep-th/0601001].

\bibitem{Ooguri:2006in} 
  H.~Ooguri and C.~Vafa,
  {\em ``On the Geometry of the String Landscape and the Swampland,''}
  Nucl.\ Phys.\ B {\bf 766}, 21 (2007)
  [hep-th/0605264].

\bibitem{Baume:2016psm} 
  F.~Baume and E.~Palti,
  {\em ``Backreacted Axion Field Ranges in String Theory,''}
  JHEP {\bf 1608}, 043 (2016)
  [arXiv:1602.06517 [hep-th]].

\bibitem{Klaewer:2016kiy} 
  D.~Klaewer and E.~Palti,
  {\em ``Super-Planckian Spatial Field Variations and Quantum Gravity,''}
  JHEP {\bf 1701}, 088 (2017)
  [arXiv:1610.00010 [hep-th]].

\bibitem{Valenzuela:2016yny} 
  I.~Valenzuela,
  {\em ``Backreaction Issues in Axion Monodromy and Minkowski 4-forms,''}
  JHEP {\bf 1706}, 098 (2017)
  [arXiv:1611.00394 [hep-th]].

\bibitem{Blumenhagen:2017cxt} 
  R.~Blumenhagen, I.~Valenzuela and F.~Wolf,
  {\em ``The Swampland Conjecture and F-term Axion Monodromy Inflation,''}
  JHEP {\bf 1707}, 145 (2017)
  [arXiv:1703.05776 [hep-th]].

\bibitem{Palti:2017elp} 
  E.~Palti,
  {\em ``The Weak Gravity Conjecture and Scalar Fields,''}
  JHEP {\bf 1708}, 034 (2017)
  [arXiv:1705.04328 [hep-th]].

\bibitem{Hebecker:2017lxm} 
  A.~Hebecker, P.~Henkenjohann and L.~T.~Witkowski,
  {\em ``Flat Monodromies and a Moduli Space Size Conjecture,''}
  JHEP {\bf 1712}, 033 (2017)
  [arXiv:1708.06761 [hep-th]].

\bibitem{Grimm:2018ohb}
  T.~W.~Grimm, E.~Palti and I.~Valenzuela,
  {\em ``Infinite Distances in Field Space and Massless Towers of States,''}
  JHEP {\bf 1808} (2018) 143
  [arXiv:1802.08264 [hep-th]].

\bibitem{Heidenreich:2018kpg} 
  B.~Heidenreich, M.~Reece and T.~Rudelius,
  {\em ``Emergence of Weak Coupling at Large Distance in Quantum Gravity,''}
  Phys.\ Rev.\ Lett.\  {\bf 121}, no. 5, 051601 (2018)
  [arXiv:1802.08698 [hep-th]].

\bibitem{Blumenhagen:2018nts} 
  R.~Blumenhagen, D.~Kl\"awer, L.~Schlechter and F.~Wolf,
  {\em ``The Refined Swampland Distance Conjecture in Calabi-Yau Moduli Spaces,''}
  JHEP {\bf 1806}, 052 (2018)
  [arXiv:1803.04989 [hep-th]].

\bibitem{Landete:2018kqf} 
  A.~Landete and G.~Shiu,
  {\em ``Mass Hierarchies and Dynamical Field Range,''}
  Phys.\ Rev.\ D {\bf 98}, no. 6, 066012 (2018)
  [arXiv:1806.01874 [hep-th]].

\bibitem{Lee:2018urn} 
  S.~J.~Lee, W.~Lerche and T.~Weigand,
  {\em ``Tensionless Strings and the Weak Gravity Conjecture,''}
  JHEP {\bf 1810}, 164 (2018)
  [arXiv:1808.05958 [hep-th]].

\bibitem{Reece:2018zvv} 
  M.~Reece,
  {\em ``Photon Masses in the Landscape and the Swampland,''}
  arXiv:1808.09966 [hep-th].
  
  \bibitem{Lee:2018spm} 
  S.~J.~Lee, W.~Lerche and T.~Weigand,
  {\em ``A Stringy Test of the Scalar Weak Gravity Conjecture,''}
  Nucl.\ Phys.\ B {\bf 938}, 321 (2019)
  [arXiv:1810.05169 [hep-th]].

 \bibitem{Ooguri:2018wrx} 
  H.~Ooguri, E.~Palti, G.~Shiu and C.~Vafa,
  {\em ``Distance and de Sitter Conjectures on the Swampland,''}
  Phys.\ Lett.\ B {\bf 788}, 180 (2019)
  [arXiv:1810.05506 [hep-th]].


\bibitem{Grimm:2018cpv}
  T.~W.~Grimm, C.~Li and E.~Palti,
  {\em ``Infinite Distance Networks in Field Space and Charge Orbits,''}
  JHEP {\bf 1903} (2019) 016
  [arXiv:1811.02571 [hep-th]].

\bibitem{Buratti:2018xjt} 
  G.~Buratti, J.~Calder\'on and A.~M.~Uranga,
  {\em ``Transplanckian Axion Monodromy !?,''}
  arXiv:1812.05016 [hep-th].

\bibitem{Hebecker:2018fln} 
  A.~Hebecker, D.~Junghans and A.~Schachner,
  {\em ``Large Field Ranges from Aligned and Misaligned Winding,''}
  JHEP {\bf 1903}, 192 (2019)
  [arXiv:1812.05626 [hep-th]].

\bibitem{Gonzalo:2018guu} 
  E.~Gonzalo, L.~E.~Ib\'a\~nez and A.~M.~Uranga,
  {\em ``Modular Symmetries and the Swampland Conjectures,''}
  arXiv:1812.06520 [hep-th].

\bibitem{Corvilain:2018lgw}
  P.~Corvilain, T.~W.~Grimm and I.~Valenzuela,
  {\em ``The Swampland Distance Conjecture for K\"ahler Moduli,''}
  arXiv:1812.07548 [hep-th].
  
\bibitem{Lee:2019tst} 
  S.~J.~Lee, W.~Lerche and T.~Weigand,
  {\em ``Modular Fluxes, Elliptic Genera, and Weak Gravity Conjectures in Four Dimensions,''}
  arXiv:1901.08065 [hep-th].
  
 \bibitem{Blumenhagen:2019qcg} 
  R.~Blumenhagen, D.~Kl\"awer and L.~Schlechter,
  {\em ``Swampland Variations on a Theme by KKLT,''}
  arXiv:1902.07724 [hep-th].

\bibitem{Joshi:2019nzi} 
  A.~Joshi and A.~Klemm,
  {\em ``Swampland Distance Conjecture for One-Parameter Calabi-Yau Threefolds,''}
  arXiv:1903.00596 [hep-th].
  
\bibitem{tensionless}
A.~Font, A.~Herr\'aez and L.~E.~Ib\'a\~nez, 
{\em ``The Swampland Distance Conjecture and Towers of Tensionless Branes,''}'
  arXiv:1904.05379 [hep-th].


\bibitem{Harlow:2015lma} 
  D.~Harlow,
  {\em ``Wormholes, Emergent Gauge Fields, and the Weak Gravity Conjecture,''}
  JHEP {\bf 1601}, 122 (2016)
  [arXiv:1510.07911 [hep-th]].

\bibitem{Heidenreich:2017sim} 
  B.~Heidenreich, M.~Reece and T.~Rudelius,
  {\em ``The Weak Gravity Conjecture and Emergence from an Ultraviolet Cutoff,''}
  Eur.\ Phys.\ J.\ C {\bf 78}, no. 4, 337 (2018)
  [arXiv:1712.01868 [hep-th]].

\bibitem{thebook}
 L.~E.~Ib\'a\~nez and A.~M.~Uranga, 
  {\it String Theory and Particle Physics. An Introduction to String Phenomenology},
  Cambridge University Press (2012). 

\bibitem{Baumann:2014nda} 
  D.~Baumann and L.~McAllister,
  {\em ``Inflation and String Theory,''}
  arXiv:1404.2601 [hep-th].

\bibitem{Kachru:2003aw} 
  S.~Kachru, R.~Kallosh, A.~D.~Linde and S.~P.~Trivedi,
  {\em ``De Sitter vacua in string theory,''}
  Phys.\ Rev.\ D {\bf 68}, 046005 (2003)
  [hep-th/0301240].

\bibitem{Balasubramanian:2005zx} 
  V.~Balasubramanian, P.~Berglund, J.~P.~Conlon and F.~Quevedo,
  {\em ``Systematics of moduli stabilisation in Calabi-Yau flux compactifications,''}
  JHEP {\bf 0503}, 007 (2005)
  [hep-th/0502058].

\bibitem{Obied:2018sgi} 
  G.~Obied, H.~Ooguri, L.~Spodyneiko and C.~Vafa,
  {\em``De Sitter Space and the Swampland,''}
  arXiv:1806.08362 [hep-th].
  
\bibitem{Garg:2018reu} 
  S.~K.~Garg and C.~Krishnan,
  {\em ``Bounds on Slow Roll and the de Sitter Swampland,''}
  arXiv:1807.05193 [hep-th].
  
 \bibitem{Rudelius:2015xta} 
  T.~Rudelius,
  {\em ``Constraints on Axion Inflation from the Weak Gravity Conjecture,''}
  JCAP {\bf 1509}, no. 09, 020 (2015)
  [arXiv:1503.00795 [hep-th]].

\bibitem{Montero:2015ofa} 
  M.~Montero, A.~M.~Uranga and I.~Valenzuela,
  {\em ``Transplanckian axions!?,''}
  JHEP {\bf 1508}, 032 (2015)
  [arXiv:1503.03886 [hep-th]].
  
 \bibitem{Brown:2015lia} 
  J.~Brown, W.~Cottrell, G.~Shiu and P.~Soler,
  {\em ``On Axionic Field Ranges, Loopholes and the Weak Gravity Conjecture,''}
  JHEP {\bf 1604}, 017 (2016)
  [arXiv:1504.00659 [hep-th]].

 \bibitem{Strominger:1995cz} 
  A.~Strominger,
  {\em ``Massless black holes and conifolds in string theory,''}
  Nucl.\ Phys.\ B {\bf 451}, 96 (1995)
  [hep-th/9504090].

\bibitem{Ooguri:1996me}
  H.~Ooguri and C.~Vafa,
  {\em``Summing up D instantons,''}
  Phys.\ Rev.\ Lett.\  {\bf 77} (1996) 3296
  [hep-th/9608079].

\bibitem{Cecotti:1988qn} 
  S.~Cecotti, S.~Ferrara and L.~Girardello,
  {\em ``Geometry of Type II Superstrings and the Moduli of Superconformal Field Theories,''}
  Int.\ J.\ Mod.\ Phys.\ A {\bf 4}, 2475 (1989).
  
  \bibitem{Ferrara:1989ik} 
  S.~Ferrara and S.~Sabharwal,
  {\em ``Quaternionic Manifolds for Type II Superstring Vacua of Calabi-Yau Spaces,''}
  Nucl.\ Phys.\ B {\bf 332}, 317 (1990).



\bibitem{deWit:1984wbb} 
  B.~de Wit and A.~Van Proeyen,
  {\em ``Potentials and Symmetries of General Gauged N=2 Supergravity: Yang-Mills Models,''}
  Nucl.\ Phys.\ B {\bf 245}, 89 (1984).
  
\bibitem{Bagger:1983tt} 
  J.~Bagger and E.~Witten,
  {\em ``Matter Couplings in N=2 Supergravity,''}
  Nucl.\ Phys.\ B {\bf 222}, 1 (1983).

\bibitem{Becker:1995kb} 
  K.~Becker, M.~Becker and A.~Strominger,
  {\em ``Five-branes, membranes and nonperturbative string theory,''}
  Nucl.\ Phys.\ B {\bf 456}, 130 (1995)
  [hep-th/9507158].

\bibitem{Marino:1999af} 
  M.~Marino, R.~Minasian, G.~W.~Moore and A.~Strominger,
  {\em ``Nonlinear instantons from supersymmetric p-branes,''}
  JHEP {\bf 0001}, 005 (2000)
  [hep-th/9911206].

\bibitem{Rocek:2005ij} 
  M.~Rocek, C.~Vafa and S.~Vandoren,
  {\em ``Hypermultiplets and topological strings,''}
  JHEP {\bf 0602}, 062 (2006)
  [hep-th/0512206].
  
  \bibitem{RoblesLlana:2006ez} 
  D.~Robles-Llana, F.~Saueressig and S.~Vandoren,
  {\em ``String loop corrected hypermultiplet moduli spaces,''}
  JHEP {\bf 0603}, 081 (2006)
  [hep-th/0602164].
  
  \bibitem{RoblesLlana:2006is} 
  D.~Robles-Llana, M.~Rocek, F.~Saueressig, U.~Theis and S.~Vandoren,
  {\em ``Nonperturbative corrections to 4D string theory effective actions from SL(2,Z) duality and supersymmetry,''}
  Phys.\ Rev.\ Lett.\  {\bf 98}, 211602 (2007)
  [hep-th/0612027].
  
  \bibitem{Neitzke:2007ke} 
  A.~Neitzke, B.~Pioline and S.~Vandoren,
  {\em ``Twistors and black holes,''}
  JHEP {\bf 0704}, 038 (2007)
  [hep-th/0701214].
  
  \bibitem{RoblesLlana:2007ae} 
  D.~Robles-Llana, F.~Saueressig, U.~Theis and S.~Vandoren,
  {\em ``Membrane instantons from mirror symmetry,''}
  Commun.\ Num.\ Theor.\ Phys.\  {\bf 1}, 681 (2007)
  [arXiv:0707.0838 [hep-th]].
  
  \bibitem{Alexandrov:2008gh} 
  S.~Alexandrov, B.~Pioline, F.~Saueressig and S.~Vandoren,
  {\em ``D-instantons and twistors,''}
  JHEP {\bf 0903}, 044 (2009)
  doi:10.1088/1126-6708/2009/03/044
  [arXiv:0812.4219 [hep-th]].
  
  \bibitem{Alexandrov:2009zh} 
  S.~Alexandrov,
  {\em ``D-instantons and twistors: Some exact results,''}
  J.\ Phys.\ A {\bf 42}, 335402 (2009)
  [arXiv:0902.2761 [hep-th]].
  
  \bibitem{Alexandrov:2010ca} 
  S.~Alexandrov, D.~Persson and B.~Pioline,
  {\em ``Fivebrane instantons, topological wave functions and hypermultiplet moduli spaces,''}
  JHEP {\bf 1103}, 111 (2011)
  [arXiv:1010.5792 [hep-th]].
  
  \bibitem{Alexandrov:2011va} 
  S.~Alexandrov,
  {\em ``Twistor Approach to String Compactifications: a Review,''}
  Phys.\ Rept.\  {\bf 522}, 1 (2013)
  doi:10.1016/j.physrep.2012.09.005
  [arXiv:1111.2892 [hep-th]].
  
 
\bibitem{Cecotti:2015wqa} 
  S.~Cecotti,
  {\em ``Supersymmetric Field Theories : Geometric Structures and Dualities,''}



\bibitem{deWit:1999fp} 
  B.~de Wit, B.~Kleijn and S.~Vandoren,
  {\em ``Superconformal hypermultiplets,''}
  Nucl.\ Phys.\ B {\bf 568}, 475 (2000)
  [hep-th/9909228].

\bibitem{deWit:2001brd} 
  B.~de Wit, M.~Rocek and S.~Vandoren,
  {\em ``Hypermultiplets, hyperKahler cones and quaternion Kahler geometry,''}
  JHEP {\bf 0102}, 039 (2001)
  [hep-th/0101161].

\bibitem{deWit:2006gn} 
  B.~de Wit and F.~Saueressig,
  {\em ``Off-shell N=2 tensor supermultiplets,''}
  JHEP {\bf 0609}, 062 (2006)
  [hep-th/0606148].

\bibitem{Saueressig:2007dr}
  F.~Saueressig and S.~Vandoren,
  {\em``Conifold singularities, resumming instantons and non-perturbative mirror symmetry,''}
  JHEP {\bf 0707} (2007) 018
  [arXiv:0704.2229 [hep-th]].

\bibitem{Collinucci:2009nv}
  A.~Collinucci, P.~Soler and A.~M.~Uranga,
  {\em ``Non-perturbative effects and wall-crossing from topological strings,''}
  JHEP {\bf 0911}, 025 (2009)
  [arXiv:0904.1133 [hep-th]].

\bibitem{Antoniadis:1997eg} 
  I.~Antoniadis, S.~Ferrara, R.~Minasian and K.~S.~Narain,
  {\em ``R**4 couplings in M and type II theories on Calabi-Yau spaces,''}
  Nucl.\ Phys.\ B {\bf 507}, 571 (1997)
  [hep-th/9707013].

\bibitem{Dvali:2007hz} 
  G.~Dvali,
  {\em ``Black Holes and Large N Species Solution to the Hierarchy Problem,''}
  Fortsch.\ Phys.\  {\bf 58}, 528 (2010)
  [arXiv:0706.2050 [hep-th]].
  
  
\bibitem{Banks:1988yz} 
  T.~Banks and L.~J.~Dixon,
  {\em ``Constraints on String Vacua with Space-Time Supersymmetry,''}
  Nucl.\ Phys.\ B {\bf 307}, 93 (1988).
  
\bibitem{Banks:2010zn} 
  T.~Banks and N.~Seiberg,
  {\em ``Symmetries and Strings in Field Theory and Gravity,''}
  Phys.\ Rev.\ D {\bf 83}, 084019 (2011)
  [arXiv:1011.5120 [hep-th]].

\bibitem{Hebecker:2017uix} 
  A.~Hebecker and P.~Soler,
  {\em ``The Weak Gravity Conjecture and the Axionic Black Hole Paradox,''}
  JHEP {\bf 1709}, 036 (2017)
  [arXiv:1702.06130 [hep-th]].


\bibitem{Hebecker:2017wsu} 
  A.~Hebecker, P.~Henkenjohann and L.~T.~Witkowski,
  {\em ``What is the Magnetic Weak Gravity Conjecture for Axions?,''}
  Fortsch.\ Phys.\  {\bf 65}, no. 3-4, 1700011 (2017)
  [arXiv:1701.06553 [hep-th]].
 
  
\bibitem{ArkaniHamed:2005yv} 
  N.~Arkani-Hamed, S.~Dimopoulos and S.~Kachru,
  {\em ``Predictive landscapes and new physics at a TeV,''}
  hep-th/0501082.
  
  \bibitem{Distler:2005hi} 
  J.~Distler and U.~Varadarajan,
  {\em ``Random polynomials and the friendly landscape,''}
  hep-th/0507090.
  
  \bibitem{Dimopoulos:2005ac} 
  S.~Dimopoulos, S.~Kachru, J.~McGreevy and J.~G.~Wacker,
  {\em ``N-flation,''}
  JCAP {\bf 0808}, 003 (2008)
  [hep-th/0507205].

\bibitem{Pioline:2009ia} 
  B.~Pioline and S.~Vandoren,
  {\em ``Large D-instanton effects in string theory,''}
  JHEP {\bf 0907}, 008 (2009)
  [arXiv:0904.2303 [hep-th]].
  
  \bibitem{Seiberg:1996ns} 
  N.~Seiberg and S.~H.~Shenker,
  {\em ``Hypermultiplet moduli space and string compactification to three-dimensions,''}
  Phys.\ Lett.\ B {\bf 388}, 521 (1996)
  [hep-th/9608086].
  

\bibitem{Kachru:1995wm} 
  S.~Kachru and C.~Vafa,
  {\em ``Exact results for N=2 compactifications of heterotic strings,''}
  Nucl.\ Phys.\ B {\bf 450}, 69 (1995)
  [hep-th/9505105].
  
  \bibitem{Aspinwall:1998bw} 
  P.~S.~Aspinwall,
  {\em ``Aspects of the hypermultiplet moduli space in string duality,''}
  JHEP {\bf 9804}, 019 (1998)
  [hep-th/9802194].

\bibitem{Louis:2011aa} 
  J.~Louis and R.~Valandro,
  {\em ``Heterotic-Type II Duality in the Hypermultiplet Sector,''}
  JHEP {\bf 1205}, 016 (2012)
  [arXiv:1112.3566 [hep-th]].



\bibitem{Huang:2006hq}
  M.~x.~Huang, A.~Klemm and S.~Quackenbush,
  {\em ``Topological string theory on compact Calabi-Yau: Modularity and boundary conditions,''}
  Lect. Notes Phys.  {\bf 757} (2009) 45
  [hep-th/0612125].
  
  \bibitem{Alexandrov:2014sya}
  S.~Alexandrov and S.~Banerjee,
  {\em``Hypermultiplet metric and D-instantons,''}
  JHEP {\bf 1502} (2015) 176
  [arXiv:1412.8182 [hep-th]].

\bibitem{Alexandrov:2009qq} 
  S.~Alexandrov and F.~Saueressig,
  {\em ``Quantum mirror symmetry and twistors,''}
  JHEP {\bf 0909}, 108 (2009)
  [arXiv:0906.3743 [hep-th]].


\end{thebibliography}
\end{document}